\RequirePackage[2020-02-02]{latexrelease}
\documentclass[aps,prd]{revtex4}

\usepackage{amsmath}
\usepackage{amssymb}
\usepackage{bm}
\usepackage{graphicx}
\usepackage{multirow}

\usepackage{textcomp}

\allowdisplaybreaks[1]

\begin{document}

\title{Dispersive analysis of neutral meson mixing}

\author{Hsiang-nan Li}
\affiliation{Institute of Physics, Academia Sinica,
Taipei, Taiwan 115, Republic of China}

\date{\today}

\begin{abstract}
We analyze the neutral meson mixing by directly solving the dispersion relation obeyed by the 
mass and width differences of the two meson mass eigenstates. We solve for the parameters 
$x$ and $y$, proportional to the mass and width differences in the charm mixing, respectively,
taking the box-diagram contributions to $x(s)$ and $y(s)$ at large mass squared $s$ of a fictitious 
$D$ meson as inputs. The SU(3) symmetry breaking is introduced through physical thresholds of 
different $D$ meson decay channels for $y(s)$. These threshold-dependent effects, acting like 
nonperturbative power corrections in QCD sum rules, stabilize the solutions of $y(s=m_D^2)$
with the $D$ meson mass $m_D$. We then calculate $x(s)$ through the dispersive integration of $y(s)$, 
and show that our predictions $x(m_D^2)\approx 0.21\%$ and $y(m_D^2)\approx 0.52\%$ are close to the 
data in both $CP$-conserving and $CP$-violating cases. It is observed 
that the channel containing di-kaon states provides the major source of SU(3) breaking, which enhances 
$x(m_D^2)$ and $y(m_D^2)$ by four orders of magnitude relative to the perturbative results. We also predict 
the coefficient ratio $q/p$ involved in the charm mixing with $|q/p|-1\approx 2\times 10^{-4}$ and 
$Arg(q/p)\approx 6\times 10^{-3}$ degrees, which can be scrutinized by precise future measurements. 
The formalism is extended to studies of the $B_{s(d)}$ meson mixing and the kaon mixing, and the
small deviations of the obtained width differences from the perturbative inputs explain why the
above mixing can be understood via short-distance dynamics. We claim that the puzzling charm mixing is
attributed to the strong Glashow-Iliopoulos-Maiani suppression on perturbative contributions, instead 
of to breakdown of the quark-hadron duality, which occurs only at 15\% level.

\end{abstract}


%
%
%

\maketitle

%
%
%
\section{INTRODUCTION}

It has been a long-standing challenge to understand the observed large $D$ meson mixing, which is 
manifested by the parameters $x$ and $y$ of order of $10^{-3}$ \cite{HFLAV:2022pwe}. The former (latter) 
is defined in terms of the mass (width) difference between the two neutral $D$ meson mass 
eigenstates. The inclusive analyses based on the heavy quark effective field theory 
\cite{Georgi:1992as,Ohl:1992sr} led to tiny $x$ and $y$ due to the strong Glashow-Iliopoulos-Maiani 
(GIM) suppression \cite{Glashow:1970gm}. The inclusion of next-to-leading-order QCD corrections yielded 
$x\sim y\simeq 6\times 10^{-7}$ \cite{Golowich:2005pt}, which fall short of the experimental
data by four orders of magnitude. It was speculated \cite{Bigi:2000wn,Falk:2001hx,Bobrowski:2010xg} that 
contributions from higher dimensional operators might circumvent the GIM suppression, and enhance $x$ and 
$y$ significantly. This speculation, requiring information on numerous nonperturbative matrix elements, has 
not been verified quantitatively. On the other hand, the exclusive analyses, where the mixing parameter $y$ 
is extracted from data of hadronic $D$ meson decays \cite{Wolfenstein:1985ft,Donoghue:1985hh,
Colangelo:1990hj,Buccella:1994nf,Buccella:1996uy,Kaeding:1995zx,Falk:2001hx,Falk:2004wg,Cheng:2010rv,
Gronau:2012kq,Jiang:2017zwr}, accounted for a half value of $y$ roughly by summing up the contributions from two-body modes \cite{Cheng:2010rv,Jiang:2017zwr}. However, it is difficult to estimate the effects
from other multi-body decays and to explain $x$ and $y$ simultaneously in this data-driven approach. 
For recent reviews on the charm mixing and related subjects, refer to 
\cite{Bhardwaj:2019vep,Umeeda:2022yao}.

The above challenge has motivated our proposal to study the $D$ meson mixing as an inverse problem, i.e., 
to solve the dispersion relation obeyed by $x(s)$ and $y(s)$ for a fictitious $D$ meson with an arbitrary 
mass squared $s$ \cite{Li:2020xrz}. The function $y(s)$ was separated into a high-mass piece and a low-mass 
piece, with the former and $x(s)$ at large $s$ being input from reliable perturbative computations of the 
box diagrams \cite{Cheng,BSS,Datta:1984jx}. The latter, treated as an unknown, was derived from the 
integral equation constructed from the dispersion relation. The unknown piece of $y(s)$ was parametrized, 
and the involved parameters were fixed by the best fit of its dispersive integral to the perturbative 
input. It turned out that many solutions of $y(s)$, corresponding to minima of the fit, were allowed as 
a consequence of the ill posed nature of an inverse problem, and those matching the data were selected.
Strictly speaking, the large $x$ and $y$ obtained in \cite{Li:2020xrz} are not 
an unambiguous prediction. The point is instead to demonstrate the existence of the nontrivial correlated 
solutions for $x$ and $y$, which, with magnitudes being much greater than from the box diagrams 
\cite{Burdman:1994th}, accommodate the large $D$ meson mixing. Nevertheless, the attempt in 
\cite{Li:2020xrz} based only on the analyticity of physical observables is novel, and has been extended to 
the constraint on the hadronic vacuum polarization contribution to the muon anomalous magnetic moment 
\cite{Li:2020fiz} and to the reformulation of QCD sum rules for determining properties of the series of
$\rho$ resonances \cite{Li:2020ejs}, glueball masses \cite{Li:2021gsx} and the pion light-cone distribution 
amplitude \cite{Li:2022qul}.

We will improve our previous work on the charm mixing \cite{Li:2020xrz} by solving the dispersion 
relation directly, which sets a stringent connection between the mixing parameters $x(s)$ and $y(s)$,
without relying on a discretionary parametrization. The advantage of the inverse matrix method developed 
in \cite{Li:2021gsx} is that a unique and stable solution can be attained before an ill posed nature 
appears. The inputs $x(s)$ and $y(s)$ at large $s$ come from the perturbative contributions, which have 
been known to explain the observed $B_{(s)}$ meson mixing satisfactorily 
\cite{Beneke:1996gn,Ciuchini:2003ww,Lenz:2006hd,Lenz:2008xt,Artuso:2015swg,Jubb:2016mvq}. It 
has been noticed that the $D$ meson mixing, strongly suppressed by the GIM mechanism, is sensitive to 
nonperturbative SU(3) symmetry breaking effects \cite{Kingsley:1975fe} characterized by the strange and 
down quark mass difference, and to Cabibbo-Kobayashi-Maskawa (CKM)-suppressed diagrams with bottom quarks 
in the loop. In our formalism the SU(3) breaking is introduced through physical thresholds for $y(s)$, 
which depend on final states of $D$ meson decays, such as $4m_K^2$ for the channel involving two strange 
quarks, $m_K$ being the kaon mass. It will be illustrated that these threshold-dependent pieces play 
a role of nonperturbative condensate, i.e., power corrections in QCD sum rules \cite{SVZ}, 
which stabilize the solutions of $y(s=m_D^2)$ solved from the above inverse problem, $m_D$ being the $D$ 
meson mass. The function $x(s)$ is then derived via the dispersive integration of the obtained $y(s)$ 
straightforwardly. We find that the results of $x(m_D^2)$ and $y(m_D^2)$ are consistent with the data 
in the $CP$-conserving case, as reasonable values for the bag parameter and the mass ratio $m_D/m_c$ 
associated with the $(S-P)(S-P)$ effective $\Delta C=2$ operator are considered, where $m_c$ is the 
charm quark mass, and $S$ ($P$) denotes the scalar (pseudoscalar) current.

It will be shown that the solution of $y(s)$ from the dispersion relation does not deviate from the 
corresponding input much for each $D$ meson decay channel actually. In other words, the quark-hadron 
duality assumed in the inclusive calculations is not broken severely for individual channel. The 
contributions from the channel containing two down quarks and the channel containing one down quark 
and one strange quark remain similar, and cancel approximately. The channel with two strange quarks, 
i.e., di-kaon states, provides the major source of the SU(3) breaking, which enhances the net 
contribution to $y(m_D^2)$ from all channels by four orders of magnitude relative to the perturbative 
one. Our observation supports the postulation \cite{Jubb:2016mvq} that a modest duality violation of 
about 20\% accounts for the huge distinction between the data and the predictions for the charm mixing 
in the inclusive analyses. Once the solutions of $x(s)$ and $y(s)$ for each channel are available, 
it is straightforward to investigate $CP$ violation in the mixing by considering
the imaginary parts of the CKM matrix elements. It will be seen 
that the resultant $x(m_D^2)$ and $y(m_D^2)$ are close to the data in the $CP$-violating case, after
the reasonable matrix element of the $(S-P)(S-P)$ operator is taken into account. At the same 
time, we predict the ratio $q/p$, where $p$ and $q$ are the coefficients relating the $D$ meson mass 
eigenstates to the flavor eigenstates through a linear combination. The prediction can be confronted by 
precise future measurements, and employed to constrain new physics models.


We then extend the formalism to studies of the $B_{s(d)}$ meson mixing and the kaon mixing. As mentioned 
before, the former can be well described in heavy quark expansion 
\cite{Beneke:1996gn,Ciuchini:2003ww,Lenz:2006hd,Lenz:2008xt}. The latter has been also explored 
intensively in perturbation theory based on the effective 
Hamiltonian, and the relevant data have been understood to some extent
\cite{Herrlich:1996vf,Buras:2010pza,Brod:2011ty}. For example, it was demonstrated \cite{Brod:2011ty} 
that short-distance contributions amount up to 89\% of the measured mass difference for the kaon mixing. 
Hence, we do not aim at precise evaluations for the above neutral meson mixing, but at a general 
picture on the mixing mechanism, and argue that they can be addressed in our framework consistently 
and systematically. As expected, the solution of the width difference is roughly equal to the 
corresponding input for each involved decay channel, similar to what is found in the $D$ meson case.
The major variation originates from the CKM matrix elements and the phase space allowed for decay 
channels. It is obvious that the GIM suppression is less effective in the $B_{s(d)}$ meson mixing with 
the different CKM factors for the up and charm quark channels. The GIM suppression is absent in the kaon 
mixing, because only the up quark channel survives the phase space constraint. We thus claim that the 
puzzling $D$ meson mixing, in contrast to the others, is attributed to the strong GIM suppression on 
the perturbative contributions in the inclusive analyses, instead of to breakdown of the quark-hadron 
duality.


The rest of the paper is organized as follows. In Sec.~II we start with the dispersion relation between the 
mixing parameters $x(s)$ and $y(s)$ for a fictitious $D$ meson, and establish the integral equation for the 
unknown function $y(s)$ that incorporates appropriate boundary conditions at physical thresholds
of involved decay channels. The SU(3) symmetry breaking effects which mimic nonperturbative 
power corrections in QCD sum rules are identified. The inverse matrix method to solve the integral 
equation is also elaborated on. The equation is solved in Sec.~III with the perturbative inputs 
from the box diagrams responsible for the charm mixing. The solution for $y(s)$ is determined, 
via whose dispersive integration the unknown function $x(s)$ is derived. The stability and reliability of 
the obtained $x(m_D^2)$ and $y(m_D^2)$ are justified. Our predictions for the relevant observables in both 
the $CP$-conserving and $CP$-violating cases are presented. We repeat the above procedures to the 
$B_{s(d)}$ meson mixing and the kaon mixing, and highlight the uniqueness of the $D$ meson mixing in 
Sec.~IV. Section V contains the conclusion and outlook.

\section{FORMALISM}

The dispersive piece $M_{12}(s)$ and the absorptive piece $\Gamma_{12}(s)$ of the analytical transition 
matrix elements, which govern the time evolution of a fictitious $D$ meson of invariant mass squared $s$, 
satisfy the dispersion relation \cite{Falk:2004wg}
\begin{eqnarray}
M_{12}(s)=\frac{1}{2\pi}\int_{4m_\pi^2}^\infty ds'\frac{\Gamma_{12}(s')}{s-s'},\label{dis}
\end{eqnarray}
where the application of the principal-value prescription to the right-hand side is implicit, and 
$4m_\pi^2$ with the pion mass $m_\pi$ is the threshold for hadronic $D$ meson decays. The mass eigenstates 
$|D_{1,2}\rangle=p|D^0\rangle\pm q|\bar D^0\rangle$ are written as the linear combinations of the 
flavor eigenstates $D^0$ and $\bar D^0$ with the coefficient ratio
\begin{eqnarray}
\frac{q}{p}=\sqrt{\frac{2M^*_{12}-i\Gamma^*_{12}}{2M_{12}-i\Gamma_{12}}}.
\end{eqnarray}
We adopt the phase convention $CP|D^0\rangle=-|\bar D^0\rangle$ for the $CP$ transformation.
The mass and width differences of the $D_{1,2}$ mesons define the mixing parameters \cite{Hagelin:1981zk} 
\begin{eqnarray}
x\equiv \frac{m_2-m_1}{\Gamma}=\frac{1}{\Gamma}{\rm Re}
\left[\frac{q}{p}(2M_{12}-i\Gamma_{12})\right],\;\;\;\;
y\equiv \frac{\Gamma_2-\Gamma_1}{2\Gamma}=-\frac{1}{\Gamma}{\rm Im}
\left[\frac{q}{p}(2M_{12}-i\Gamma_{12})\right],
\label{3}
\end{eqnarray}
with the total decay width $\Gamma$, which reduce to
\begin{eqnarray}
x=\frac{2M_{12}}{\Gamma},\;\;\;\;
y=\frac{\Gamma_{12}}{\Gamma},
\label{xyd}
\end{eqnarray} 
in the $CP$-conserving case. The masses of the other quarks maintain their physical values, so the 
fictitious $D$ meson decays into the allowed final states, as its mass crosses each threshold.


We decompose the absorptive piece into 
\begin{eqnarray}
\Gamma_{12}(s)=\sum_{i,j}\lambda_{i}\lambda_{j} \Gamma_{ij}(s),\label{dec}
\end{eqnarray} 
with the internal quarks $i,j=d,s,b$, and $\lambda_k\equiv V_{ck}V^*_{uk}$, $k=d,s,b$, being the 
products of the CKM matrix elements. The component $\Gamma_{ij}(s)$, calculable perturbatively at large 
$s$, approaches the box-diagram contribution
\begin{eqnarray}
\Gamma_{ij}^{\rm box}(s)=\frac{G_F^2 f_D^2m_W^3 B_D}{12\pi^2}A_{ij}^{\rm box}(s),
\label{CKM}
\end{eqnarray}
where $G_F$ is the Fermi constant, $f_D$ is the $D$ meson decay constant, $m_W$ is
the $W$ boson mass and $B_D$ is the bag parameter. The perturbative function $A_{ij}^{\rm box}$ combines 
the results from the $(V-A)(V-A)$ and $(S-P)(S-P)$ operators in the effective weak Hamiltonian \cite{BSS},
\begin{eqnarray}
A_{ij}^{\rm box}(s)&=&\frac{\pi}{2x_D^{3/2}}
\frac{\sqrt{x_D^2-2x_D(x_i+x_j)+(x_i-x_j)^2}}{(1-x_i)(1-x_j)}\nonumber\\
& &\times\left\{\left(1+\frac{x_ix_j}{4}\right)
[3x_D^2-x_D(x_i+x_j)-2(x_i-x_j)^2]+2x_D(x_i+x_j)(x_i+x_j-x_D)\right\},
\label{aij}
\end{eqnarray}
which is symmetric under the exchange of the subscripts $i$ and $j$, i.e., 
$A_{ij}^{\rm box}(s)=A_{ji}^{\rm box}(s)$. We have flipped the sign of the formula in \cite{BSS} 
to match the convention in Eq.~(\ref{3}). In the above expression the variables are defined as 
$x_i=m_i^2/m_W^2$, $m_i$ being the mass of the quark $i$, and $x_D=s/m_W^2$. Note that $A_{ij}^{\rm box}$
contribute up to $A_{dd}^{\rm box}$ ($A_{ds}^{\rm box}$, $A_{ss}^{\rm box}$, $A_{db}^{\rm box}$, 
$A_{sb}^{\rm box}$, $A_{bb}^{\rm box}$) allowed in the range 
$s < (m_d+m_s)^2$ [$(m_d+m_s)^2\le s < 4m_s^2$, 
$4m_s^2\le s < (m_d+m_b)^2$, $(m_d+m_b)^2\leq s<(m_s+m_b)^2$, $(m_s+m_b)^2\leq s<4m_b^2$, $4m_b^2\le s$].

\begin{figure}
\begin{center}
\includegraphics[scale=0.3]{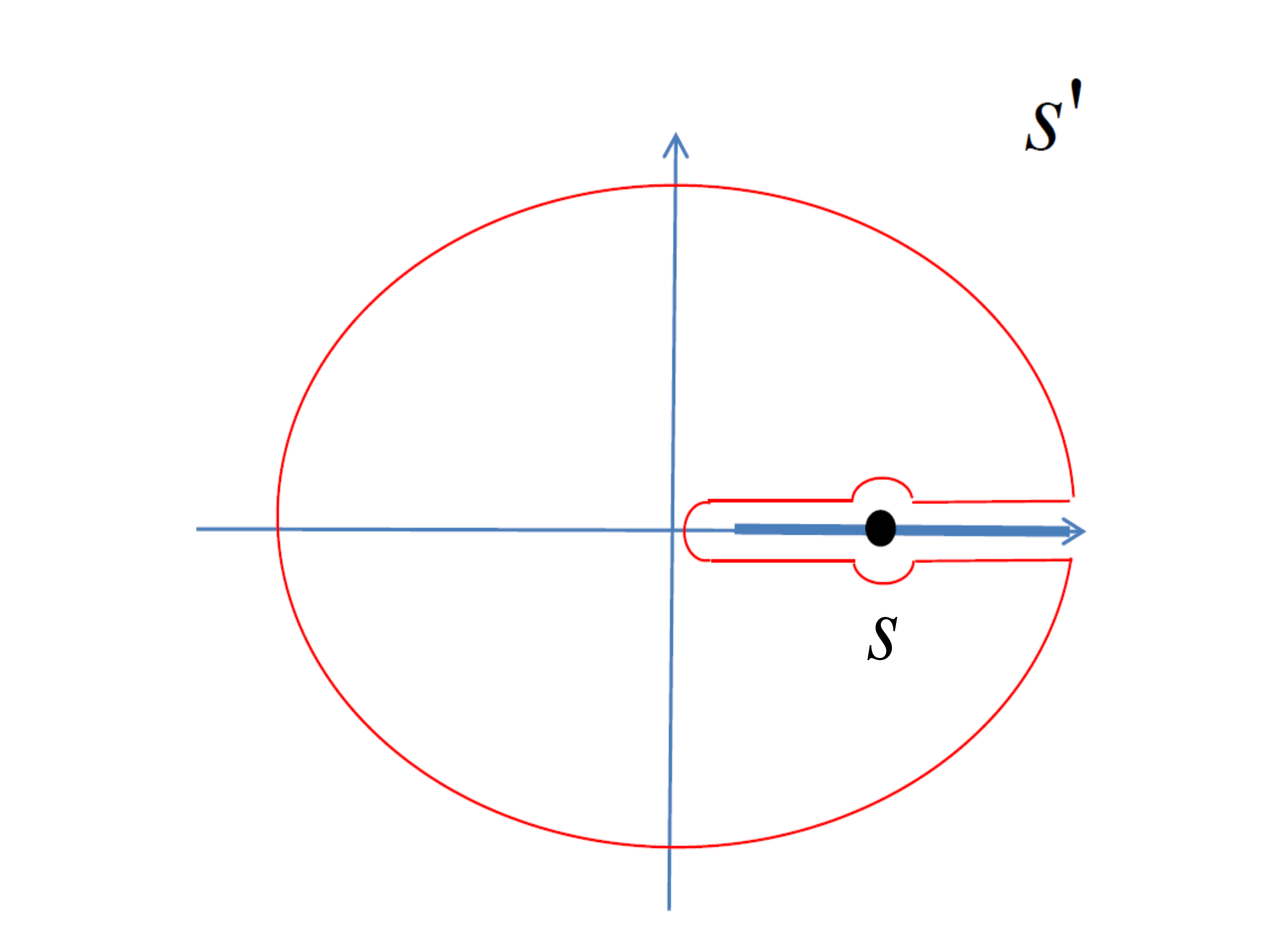}
\caption{\label{fig0}
Contour considered in Eq.~(\ref{con}), where the thick line represents the branch cut.}
\end{center}
\end{figure}

Similarly, we decompose the dispersive piece into $M_{12}(s)=\sum_{i,j}\lambda_{i}\lambda_{j} M_{ij}(s)$. 
In principle, the dispersion relation, as a result of QCD dynamics which has nothing to do with the CKM 
factors, holds for each pair of the components $M_{ij}(s)$ and $\Gamma_{ij}(s)$. This fact has been noticed 
in \cite{Li:2020xrz}, but was not implemented in the preliminary attempt there. Though $\Gamma_{12}(s)$ 
deceases fast enough at large $s$ \cite{Li:2020xrz}, so that the dispersive integral on the right hand side 
of Eq.~(\ref{dis}) converges, each component $\Gamma_{ij}(s)$ grows like $s^{3/2}$ as indicated in 
Eq.~(\ref{aij}). These divergent behaviors cancel in the sum in Eq.~(\ref{dec}), when the unitarity 
of the CKM factors is imposed. We thus reformulate the dispersion relation, starting with the contour 
integral for the analytical function $\Pi_{ij}(s)=M_{ij}(s)-i\Gamma_{ij}(s)/2$,
\begin{eqnarray}
\frac{1}{2\pi i}\oint ds'\frac{\Pi_{ij}(s')}{s-s'}=0.\label{con}
\end{eqnarray}
The contour in Eq.~(\ref{con}) consists of two pieces of horizontal lines above and below the branch cut 
along the positive real axis on the complex $s'$ plane, a circle of small radius $r$ around the pole 
$s'=s$ located on the positive real axis, and a circle $C_R$ of large radius $R$ as depicted in 
Fig.~\ref{fig0}. The integral vanishes, since 
there is no pole in the contour, which encloses only unphysical regions. The contribution along the small 
clockwise circle yields $M_{ij}$, and that from the two pieces of horizontal lines leads to 
the dispersive integral of $\Gamma_{ij}$. Equation~(\ref{con}) then gives
\begin{eqnarray}
M_{ij}(s)=\frac{1}{2\pi}\int_{m_{IJ}}^R ds'\frac{\Gamma_{ij}(s')}{s-s'}
+\frac{1}{2\pi i}\int_{C_R} ds'\frac{\Pi_{ij}^{\rm box}(s')}{s-s'},\label{ij}
\end{eqnarray}
where $m_{IJ}$ represents the threshold mass squared of the hadronic states contributing to
$\Gamma_{ij}$, such as $m_{\pi \pi}=4m_\pi^2$, $m_{\pi K}=(m_\pi+m_K)^2$, $m_{KK}=4m_K^2$, 
$m_{\pi B}=(m_\pi+m_B)^2$, ... with the $B$ meson mass $m_B$. The unknown function $\Gamma_{ij}(s)$, 
containing nonperturbative dynamics from the low $s$ region, will be solved from the dispersion relation. 
The integrand $\Pi_{ij}$, taking values along the large clockwise circle $C_R$, can be safely replaced 
by the perturbative expression $\Pi_{ij}^{\rm box}$ from the box-diagram computation.



The dispersive piece $M_{ij}^{\rm box}(s)$ and the absorptive piece $\Gamma_{ij}^{\rm box}(s)$  associated 
with the box diagrams respect the dispersion relation apparently,
\begin{eqnarray}
M_{ij}^{\rm box}(s)=\frac{1}{2\pi}\int_{m_{ij}}^R ds'\frac{\Gamma_{ij}^{\rm box}(s')}{s-s'}
+\frac{1}{2\pi i}\int_{C_R} ds'\frac{\Pi_{ij}^{\rm box}(s')}{s-s'},\label{ope}
\end{eqnarray}
where $m_{ij}$ is the threshold mass squared of the quark states contributing to $\Gamma_{ij}^{\rm box}$, 
such as $m_{dd}=4m_d^2$, $m_{ds}=(m_d+m_s)^2$, $m_{ss}=4m_s^2$, $m_{db}=(m_d+m_b)^2$, ....
Because a heavy neutral meson mixing can be well described by perturbative contributions, 
we approximate $M_{ij}(s)$ by $M_{ij}^{\rm box}(s)$, i.e., equate Eqs.(\ref{ij}) and (\ref{ope}) at large 
enough $s$, arriving at 
\begin{eqnarray}
\int_{m_{IJ}}^R ds'\frac{\Gamma_{ij}(s')}{s-s'}=
\int_{m_{ij}}^R ds'\frac{\Gamma_{ij}^{\rm box}(s')}{s-s'},\label{ij2}
\end{eqnarray}
where the box-diagram contributions from the large circle $C_R$ on the two sides have canceled.

It has been emphasized \cite{Li:2021gsx} that the boundary condition of an unknown function is crucial 
for the determination of its solutions from a dispersion relation. As $s$ is near a threshold, the 
fictitious $D$ meson decay is dominated by a single mode $D\to PP$, with $P$ representing a light 
pseudoscalar meson of mass $m_P$. For instance, the components $\Gamma_{dd}$, $\Gamma_{ds}$, and 
$\Gamma_{ss}$ are dominated by $D\to \pi\pi$, $\pi K$, and $K K$, respectively, when the fictitious 
$D$ meson mass approaches the corresponding thresholds from above. The $D\to PP$ decay width is 
proportional to $p_c|{\cal M}|^2/s$, with $p_c$ being the center-of-mass momentum of the pseudoscalar 
meson, and the amplitude ${\cal M}\propto s-m_P^2$ in the naive factorization assumption. It is then 
easy to acquire the power-law behaviors $p_c\sim O(m_P)$, ${\cal M}\sim O(m_P^2)$ and 
$\Gamma_{ij}\sim O(m_P^3)$ around the threshold $s\sim O(m_P^2)$. The naive factorization assumption 
may not be reliable in the above low-mass regions, but it is interesting to note that the obtained 
observation is the same as deduced from the $K\to\pi\pi$ amplitude in chiral perturbation 
theory \cite{Falk:2004wg}. Certainly, the above argument does not apply to the boundary conditions of the 
components $\Gamma_{(d,s,b)b}$, to which the states containing heavy $B$ mesons contribute: the similar 
reasoning leads to $\Gamma_{db}\sim \Gamma_{sb} \sim O(m_P)$ at $s\sim (m_B+m_P)^2$. 
Nevertheless, we will assume the same threshold behaviors for the derivations of $\Gamma_{(d,s,b)b}$, 
since their contributions to the $D$ meson mixing are negligible owing to the strong suppression by the 
CKM factors as explicitly verified in the next section. 

Following the procedure in \cite{Li:2021gsx}, we introduce a subtracted unknown function 
$\Delta {\Gamma}_{ij}$, which is related to the original $\Gamma_{ij}$ via
\begin{eqnarray}
\Delta{\Gamma}_{ij}(s,\Lambda)={\Gamma}_{ij}(s)-\Gamma_{ij}^{\rm box}(s)
\{1-\exp[-(s-m_{IJ})^2/\Lambda^2]\}.
\label{sub}
\end{eqnarray}
The scale $\Lambda$ characterizes the order of $s$, at which ${\Gamma}_{ij}(s)$ transits to the 
perturbative expression $\Gamma_{ij}^{\rm box}(s)$. The subtraction term in Eq.~(\ref{sub}) vanishes 
like $(s-m_{IJ})^2\sim O(m_{IJ}^2)$ near the threshold $s\sim O(m_{IJ})$, because 
$\Gamma_{ij}^{\rm box}(m_{IJ})$ with $m_{IJ}>m_{ij}$ is finite as implied by Eq.~(\ref{aij}). Namely, 
$\Delta{\Gamma}_{ij}(s,\Lambda)$ exhibits the low-mass behavior the same as 
${\Gamma}_{ij}(s)\sim O(m_{IJ}^{3/2})$. We have tested other choices of the subtraction function, 
like $1-\exp[-(s-m_{IJ})^3/\Lambda^3]$, which diminishes more rapidly as $s\to m_{IJ}$ and does not 
modify the low-mass behavior of ${\Gamma}_{ij}(s)$ either, and made sure that our solutions for the 
mixing parameters alter by only few percent. The function $1-\exp[-(s-m_{IJ})^2/\Lambda^2]$ in 
Eq.~(\ref{sub}) approaches unity, i.e., $\Delta{\Gamma}_{ij}(s,\Lambda)$ vanishes only at large 
$s\gg\Lambda$. In other words, the quark-hadron duality is not postulated at any finite $s$ in our 
formalism.



The subtraction term in Eq.~(\ref{sub}) can be regarded as an ultraviolet regulator for a 
dispersive integral mentioned in \cite{Forkel:2003mk}. The dispersive integral, formulated with the 
subtracted unknown function, then converges, and Eq.~(\ref{ij2}) can be rewritten as
\begin{eqnarray}
\int_{m_{IJ}}^\infty ds'\frac{\Delta \Gamma_{ij}(s',\Lambda)}{s-s'}=
\int_{m_{IJ}}^\infty ds'\frac{\Gamma_{ij}^{\rm box}(s')\exp[-(s'-m_{IJ})^2/\Lambda^2]}{s-s'}
+\int_{m_{ij}}^{m_{IJ}} ds'\frac{\Gamma_{ij}^{\rm box}(s')}{s-s'},\label{ij24}
\end{eqnarray}
where the upper bounds $R$ have been pushed to infinity due to the finiteness of the integrals. 
Note that an emitted $W$ boson can become real when $s$ is large enough, and the expression of
$\Gamma_{ij}^{\rm box}(s)$ should be modified. As observed in the next section, the scale $\Lambda$ takes 
values of order of few GeV$^2$, so the concerned high-mass region, greatly suppressed by the exponential 
factor $\exp[-(s'-m_{IJ})^2/\Lambda^2]$, is not important. Strictly speaking, the decay constant 
$f_D$ and the bag parameter $B_D$ depend on the fictitious $D$ meson mass. However, the decay constants 
of the physical pseudoscalar mesons do not vary much in the low $s$ region, ranging from 
$m_\pi^2\approx 0.02$ GeV$^2$ to $m_{B_s}^2\approx 29$ GeV$^2$, to which Eq.~(\ref{ij24}) is relevant. 
That is, the value of $f_D$ does not matter to the explanation of the $10^4$ enhancement factor. 
The bag parameters fluctuate only a bit in the range of $s$ from $m_K^2\approx 0.25$ GeV$^2$ to 
$m_{B_s}^2\approx 29$ GeV$^2$ as shown in the lattice calculations 
\cite{Hashimoto:1999ck,FermilabLattice:2016ipl,Carrasco:2015pra,Dowdall:2019bea}. Hence,
it is numerically appropriate to treat both $f_D$ and $B_D$
as constants in the dominant $s'$ region for Eq.~(\ref{ij24}).

We then remove the common constant prefactors on the two sides of Eq.~(\ref{ij24}), and replace 
$\Delta \Gamma_{ij}(s,\Lambda)$ [$\Gamma_{ij}^{\rm box}(s)$] by the function $\Delta A_{ij}(s,\Lambda)$ 
[$A_{ij}^{\rm box}(s)$] according to Eq.~(\ref{CKM}). 
Since $\Delta A_{ij}(s,\Lambda)$ is a dimensionless quantity, it can be cast into the form 
$\Delta A_{ij}(s/\Lambda)$. Other ratios like $s/m_{IJ}$ can be reexpressed as 
$(s/\Lambda)(\Lambda/m_{IJ})$, 
so $s/\Lambda$ is the only variable of $\Delta A_{ij}$. Equation~(\ref{ij24}) becomes, under the 
substitution $s'\to s'+m_{IJ}$ and the variable changes $s-m_{IJ}=u\Lambda$, $s'=v\Lambda$, 
$m_{ij}=r_{ij}\Lambda$ and $m_{IJ}=r_{IJ}\Lambda$,
\begin{eqnarray}
\int_{0}^\infty dv\frac{\Delta A_{ij}(v)}{u-v}=
\int_{0}^\infty dv\frac{A_{ij}^{\rm box}(v\Lambda+m_{IJ})e^{-v^2}}{u-v}
+\int_{r_{ij}-r_{IJ}}^0 dv\frac{A_{ij}^{\rm box}(v\Lambda+m_{IJ})}{u-v}.\label{r21}
\end{eqnarray}
The lower bound of the second term on the right-hand side represents the sources of
nonperturbative dynamics with $m_{IJ}\not= m_{ij}$, and of the SU(3) symmetry breaking 
with the dependence of $m_{IJ}$ on the hadronic states labelled by $IJ$. 
The solutions for the mixing parameters, as physical observables, should be insensitive to the transition 
scale $\Lambda$, which is introduced through the ultraviolet regulation for the dispersive integrals.
It will be elaborated that the second term on the right-hand side of Eq.~(\ref{r21}) plays the role of 
nonperturbative condensate, i.e., power corrections in QCD sum rules, which stabilize the solutions with 
respect to the variation of $\Lambda$. When $\Lambda$ increases, the magnitude of the first integral on 
the right-hand side grows, for $A_{ij}^{\rm box}$ behaves monotonically with $s$. On the contrary, the 
second integral picks up values of $A_{ij}^{\rm box}$ at lower $s$ specified by the integration interval, 
where $A_{ij}^{\rm box}$ changes slowly. The shrinking of the integration interval with $\Lambda$ yields 
stronger reduction, such that the magnitude of the second integral decreases. It is possible that the 
changes of the two terms compensate each other, and stable solutions may exist in a window of $\Lambda$, 
which are then identified as our results for the mixing parameters. The numerical analysis to be performed 
in the next section does reveal the stability of the solutions.

Viewing the boundary condition of $\Delta A_{ij}(v)\sim v^{3/2}$ at $v\to 0$, 
we expand it in terms of the generalized Laguerre polynomials $L_n^{(\alpha)}(v)$ 
for the parameter $\alpha=3/2$,
\begin{eqnarray}
\Delta A_{ij}(v)=\sum_{n=1}^N a^{(ij)}_{n}v^\alpha e^{-v}L_{n-1}^{(\alpha)}(v),\label{r0}
\end{eqnarray}
up to degree $N-1$ with the unknown coefficients $a^{(ij)}_{n}$. The generalized Laguerre polynomials 
obey the orthogonality
\begin{eqnarray}
\int_0^\infty v^\alpha e^{-v}L_m^{(\alpha)}(v)L_n^{(\alpha)}(v)dv=\frac{\Gamma(m+\alpha+1)}{m!}\delta_{mn}.
\label{or0}
\end{eqnarray}  
The number of polynomials $N$ should be as large as possible, such that Eq.~(\ref{r0}) best describes the 
subtracted unknown function, but cannot be too large in order to avoid the appearance of an ill posed 
nature. Because $\Delta A_{ij}(v)$ decreases quickly enough with $v$, as designed in Eq.~(\ref{sub}), 
the major contribution to its integral arises from a finite range of $v$. It is then justified to expand 
the integral on the left-hand side of Eq.~(\ref{r21}) into a series in $1/u$ up to the power $N$ for a 
sufficiently large $|u|$ by inserting
\begin{eqnarray}
\frac{1}{u-v}=\sum_{m=1}^N \frac{v^{m-1}}{u^m}.
\label{ep0}
\end{eqnarray}
The right-hand side of Eq.~(\ref{r21}) can be expanded into a power series in $1/u$:
the exponential factor $e^{-v^2}$ in the first integral diminishes the contribution from large $v$, and 
$v$ is restricted in a finite interval in the second integral.

Substituting Eqs.~(\ref{r0}) and (\ref{ep0}) into Eq.~(\ref{r21}), and equating the coefficients 
of $1/u^m$ in the power series on the two sides of Eq.~(\ref{r21}), we construct the matrix equation 
$Ua^{(ij)}=b^{(ij)}$ with the matrix elements
\begin{eqnarray}
U_{mn}&=&\int_{0}^\infty dv v^{m-1+\alpha}e^{-v}L_{n-1}^{(\alpha)}(v),\label{mm2}
\end{eqnarray}
where $m$ and $n$ run from 1 to $N$. We have $U_{mn}=0$ actually for $n>m$ with the orthogonality
condition in Eq.~(\ref{or0}). The vector 
\begin{eqnarray}
a^{(ij)}=(a^{(ij)}_{1}, a^{(ij)}_{2},\cdots,a^{(ij)}_{N}),\label{an1}
\end{eqnarray}
collects the unknowns. The power expansion on the right-hand side of Eq.~(\ref{r21}) gives the coefficient 
$b^{(ij)}_m$ of the term $1/u^m$, i.e., the $m$th element of the input vector $b^{(ij)}$,
\begin{eqnarray}
b^{(ij)}_m =\int_{0}^\infty dvv^{m-1} A_{ij}^{\rm box}(v\Lambda+m_{IJ})e^{-v^2}
+\int_{r_{ij}-r_{IJ}}^0 dvv^{m-1}A_{ij}^{\rm box}(v\Lambda+m_{IJ}).\label{opei}
\end{eqnarray}
One can then solve for the vector $a^{(ij)}$ through $a^{(ij)}=U^{-1}b^{(ij)}$ by applying the 
inverse matrix $U^{-1}$. The existence of $U^{-1}$ implies the uniqueness of the solution for $a^{(ij)}$.  
An inverse problem is usually ill posed; namely, some elements of $U^{-1}$ rise fast with its dimension.
Nevertheless, the convergence of Eq.~(\ref{r0}) can be achieved at a finite $N$, before $U^{-1}$ goes out 
of control. The difference between an obtained solution and a true one produces a correction to 
Eq.~(\ref{r21}) only at power $1/u^{N+1}$, and the coefficients $a_n^{(ij)}$ built up previously 
are not altered by the inclusion of an additional higher-degree polynomial into the expansion 
in Eq.~(\ref{r0}), because of the orthogonality condition in Eq.~(\ref{or0}). The convergence of solutions 
in the polynomial expansion and their insensitivity to $\Lambda$ will validate our approach, which is 
thus free of tunable parameters.  

We get ${A}_{ij}(s)$ from $\Delta{A}_{ij}(s,\Lambda)$ by adding back the subtraction term, 
and the solution
\begin{eqnarray}
y(s)=\frac{G_F^2 f_D^2m_W^3 B_D}{12\pi^2\Gamma}\sum_{i,j}\lambda_{i}\lambda_{j} 
\left\{\Delta A_{ij}(s,\Lambda)
+A_{ij}^{\rm box}(s)\left[1-e^{-(s-m_{IJ})^2/\Lambda^2}\right]\right\},\label{y1}
\end{eqnarray}
in which only the components with $m_{IJ}<m_D^2$ contribute to the physical value $y(m_D^2)$. In principle, 
one can evaluate $x(s)$ by inserting Eq.~(\ref{y1}) into Eq.~(\ref{dis}). Note that the integration of the 
subtraction term to $s\to\infty$ in Eq.~(\ref{dis}) develops divergences, which ought to cancel in the 
summation over $i,j$. This delicate cancellation renders numerical outcomes unstable. A trick is to 
utilize the facts that the contributions to $x(s)$ and $y(s)$ from the box diagrams satisfy the dispersion 
relation in Eq.~(\ref{ope}), and that they are four orders of magnitude smaller than our solutions as seen 
later. We then have
\begin{eqnarray}
x(s)=\frac{G_F^2 f_D^2m_W^3 B_D}{12\pi^3\Gamma}\sum_{i,j}\lambda_{i}\lambda_{j} 
\left\{\int_{m_{IJ}}^\infty \frac{ds'}{s-s'}\left[\Delta A_{ij}(s',\Lambda)
-A_{ij}^{\rm box}(s')e^{-(s-m_{IJ})^2/\Lambda^2}\right]
+\int_{m_{IJ}}^{m_{ij}}\frac{ds'}{s-s'} A_{ij}^{\rm box}(s')\right\},\label{xy}
\end{eqnarray}
where the integrals of $A_{ij}^{\rm box}(s')$ in the interval $[m_{ij},\infty)$ have been dropped
in the light of the above argument. It is obvious that each term on the right-hand side of Eq.~(\ref{xy})
is convergent. Our formalism can be extended to investigations of other neutral meson mixing 
straightforwardly with appropriate replacements of quark flavors, hadronic states, and the CKM matrix 
elements.


\section{$D$ MESON MIXING}

We first conduct the numerical analysis of the $D$ meson mixing using the method developed in the previous 
section with the Fermi constant $G_F=1.1663788\times 10^{-5}$ GeV$^{-2}$, the  
decay constant $f_D=0.213$ GeV, the $D$ meson decay width $\Gamma=1.60 \times 10^{-12}$ GeV 
(corresponding to the lifetime $\tau = 410.3\times 10^{-15}$ s), the masses 
$m_D = 1.865$ GeV, $m_d = 0.005$ GeV, $m_s = 0.093$ GeV, $m_b = 4.8$ GeV and $m_W = 80.377$ GeV,
the Wolfenstein parameters $\lambda = 0.225$, $A= 0.826$, $\bar\rho = 0.159$ and $\bar\eta = 0.348$ 
for the CKM matrix elements \cite{PDG}, and the typical bag parameter $B_D \approx 1$.
The unitarity of the CKM matrix turns Eq.~(\ref{dec}) for $s=m_D^2$ into
\begin{eqnarray}
\Gamma_{12}(m_D^2)=
\lambda_{s}^2[\Gamma_{dd}(m_D^2)-2\Gamma_{ds}(m_D^2)+\Gamma_{ss}(m_D^2)]
+2\lambda_s\lambda_b[\Gamma_{dd}(m_D^2)-\Gamma_{ds}(m_D^2)]
+\lambda_{b}^2\Gamma_{dd}(m_D^2),\label{g12}
\end{eqnarray}
which indicates clearly that the charm mixing is sensitive to the flavor symmetry breaking.
Substituting Eq.~(\ref{CKM}) from the box diagrams into the above expression, we find in the 
$CP$-conserving case, where only the real part of the CKM matrix element $V_{ub}$ is considered, that the 
$\lambda_s\lambda_b$ piece is positive with its magnitude being larger than of the negative $\lambda_s^2$ 
piece. The $\lambda_b^2$ piece, being of order of $10^{-8}$, is negligible compared with the first two, 
which are of order of $10^{-7}$. In total, the box diagrams contribute $3.7\times 10^{-7}$ to the 
parameter $y$ for the $D$ meson mixing, lower than the measured value by four orders of magnitude.

\begin{figure}
\begin{center}
\includegraphics[scale=0.55]{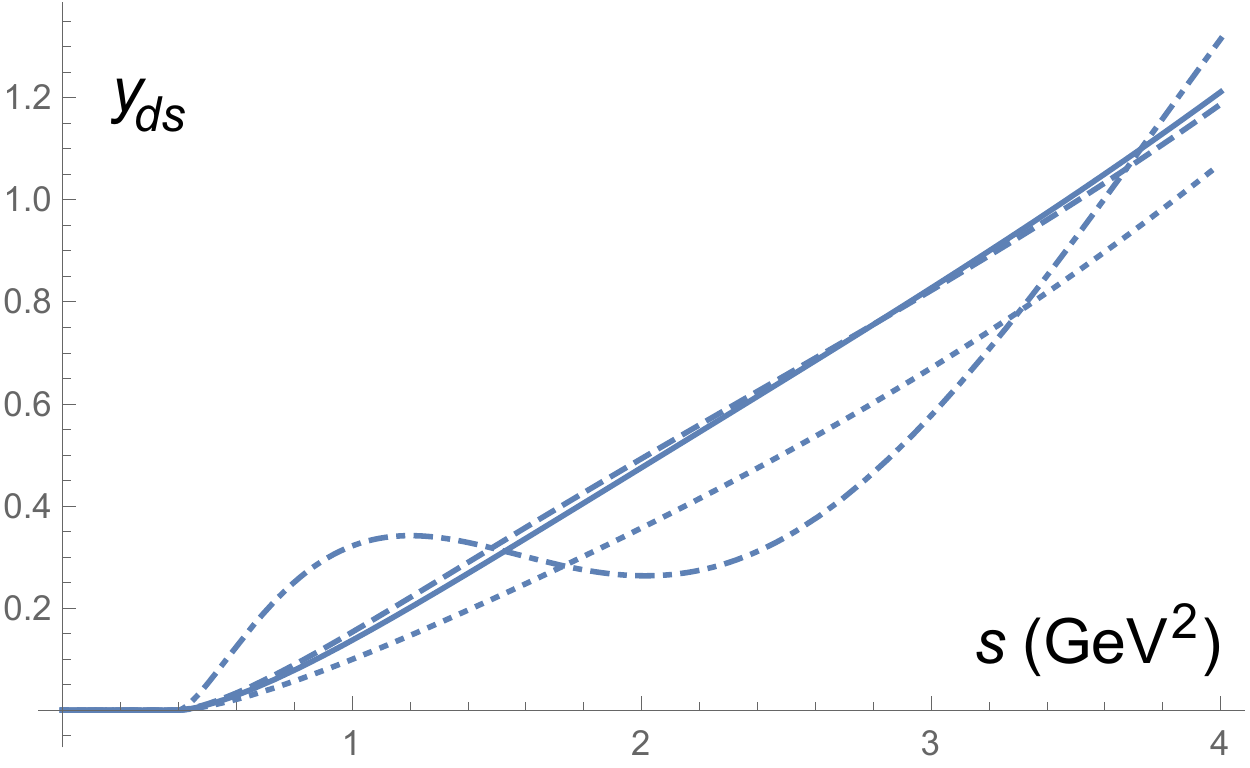}
\caption{\label{fig1}
Dependencies of $y_{ds}(s)\equiv \Gamma_{ds}(s)/\Gamma$ on $s$ for $N=3$ (dotted line),
$N=8$ (dashed line), $N=13$ (solid line) and $N=23$ (dot-dashed line) with $\Lambda=5$ GeV$^2$.}
\end{center}
\end{figure}

We solve for the component $\Gamma_{ds}(s)$ in the decomposition of $\Gamma_{12}(s)$ in Eq.~(\ref{dec}) 
as a demonstration, computing the matrix $U$ in Eq.~(\ref{mm2}) and the input vector $b^{(ds)}$ in 
Eq.~(\ref{opei}) for a given transition scale $\Lambda$, and deriving the unknown vector 
$a^{(ds)}=U^{-1}b^{(ds)}$. The dimension $N$ of the matrix $U$ is increased one by one to search for 
a convergent expansion in Eq.~(\ref{r0}). When the convergence is attained, the solutions of $a^{(ds)}$ 
and of $\Delta A_{ds}(s,\Lambda)$ become stable with respect to the variation of $N$, which are then 
selected to form the solutions of $\Gamma_{ds}(s)$ in Eq.~(\ref{sub}). We list 
$a_n^{(ds)}$ for $\Lambda=5$ GeV$^2$ up to $n=23$ below,
\begin{eqnarray}
& &10^5\times(a_1^{(ds)},a_2^{(ds)},a_3^{(ds)},\cdots,a_{12}^{(ds)},a_{13}^{(ds)},a_{14}^{(ds)},\cdots,
a_{22}^{(ds)},a_{23}^{(ds)})\nonumber\\
&=&(4.04, 2.47, 1.45 ,\cdots,
-2.08\times 10^{-2}, -4.59\times 10^{-3}, 9.25\times 10^{-3},\cdots,
7.49\times 10^{-2}, 1.04),\label{an}
\end{eqnarray}
whose magnitudes keep decreasing till $n=13$, then increase with $n$, and $a_{23}^{(ds)}$ becomes as 
large as the first few coefficients. The small ratio $|a_{13}^{(ds)}/a_1^{(ds)}|\approx 10^{-3}$ marks a
satisfactory convergence of the series up to $n=13$, and the ill posed nature emerges gradually afterwards.
We display the functions $y_{ds}(s)\equiv \Gamma_{ds}(s)/\Gamma$ corresponding to $N=3$, 8, 13 and 23 for 
the expansion in Eq.~(\ref{r0}) in Fig.~\ref{fig1}. The dependencies on $s$ match the pattern of 
Eq.~(\ref{an}): the curve of $N=3$ differs from those of $N=8$ and $N=13$, which coincide with each other 
approximately. In fact, the curves for $N$ around 13, including $N=11$-15, overlap perfectly, confirming 
the convergence of the expansion in $N$. The curve of $N=23$ with obvious oscillations signals that the 
matrix elements of $U^{-1}$ have gone out of control. The above examination suggests that $N=13$ is the 
optimal choice, and the corresponding $y_{ds}(s)$ is the solution for the given $\Lambda=5$ GeV$^2$.

\begin{figure}
\begin{center}
\includegraphics[scale=0.4]{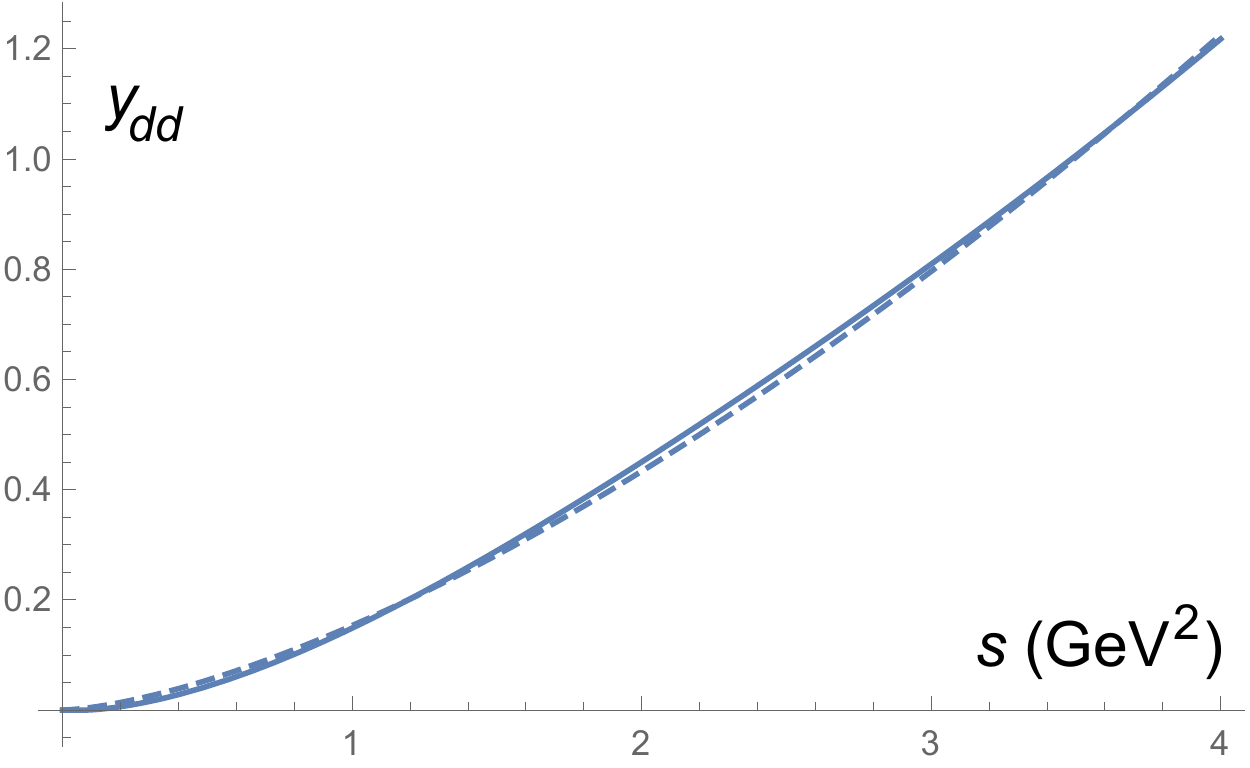}
\includegraphics[scale=0.4]{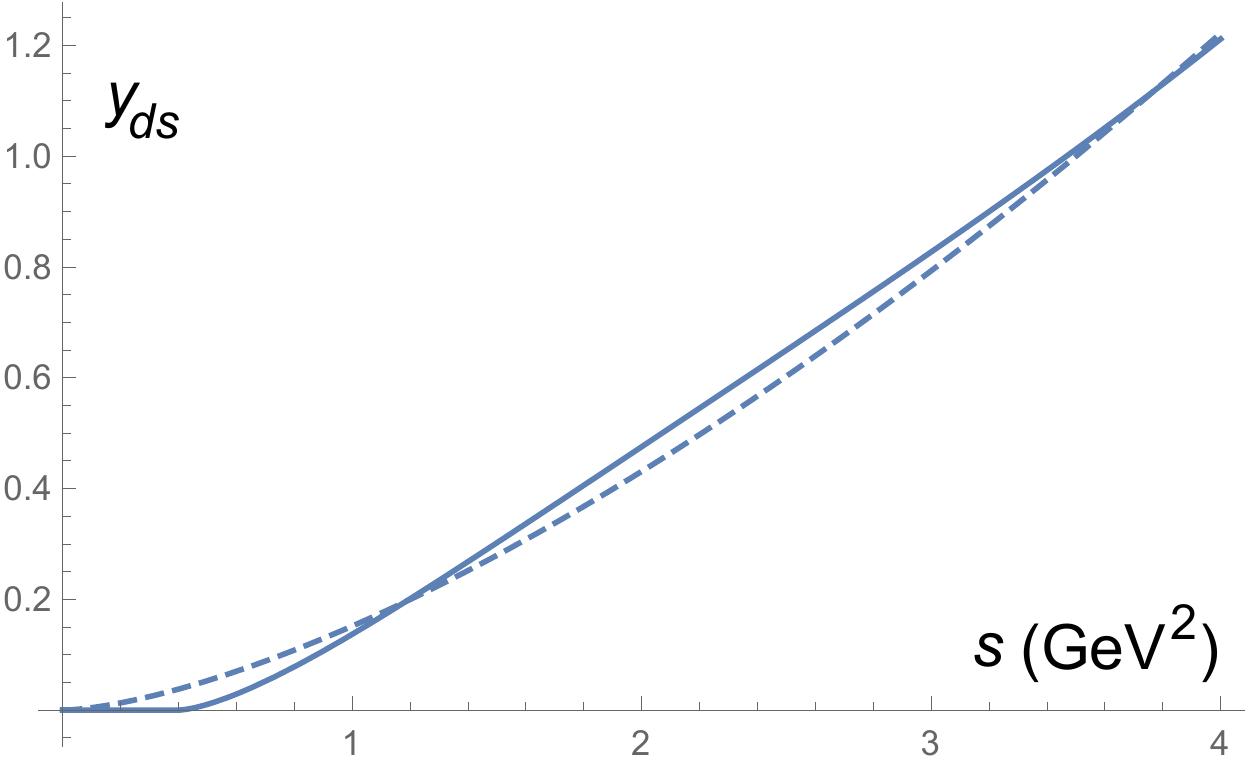}
\includegraphics[scale=0.4]{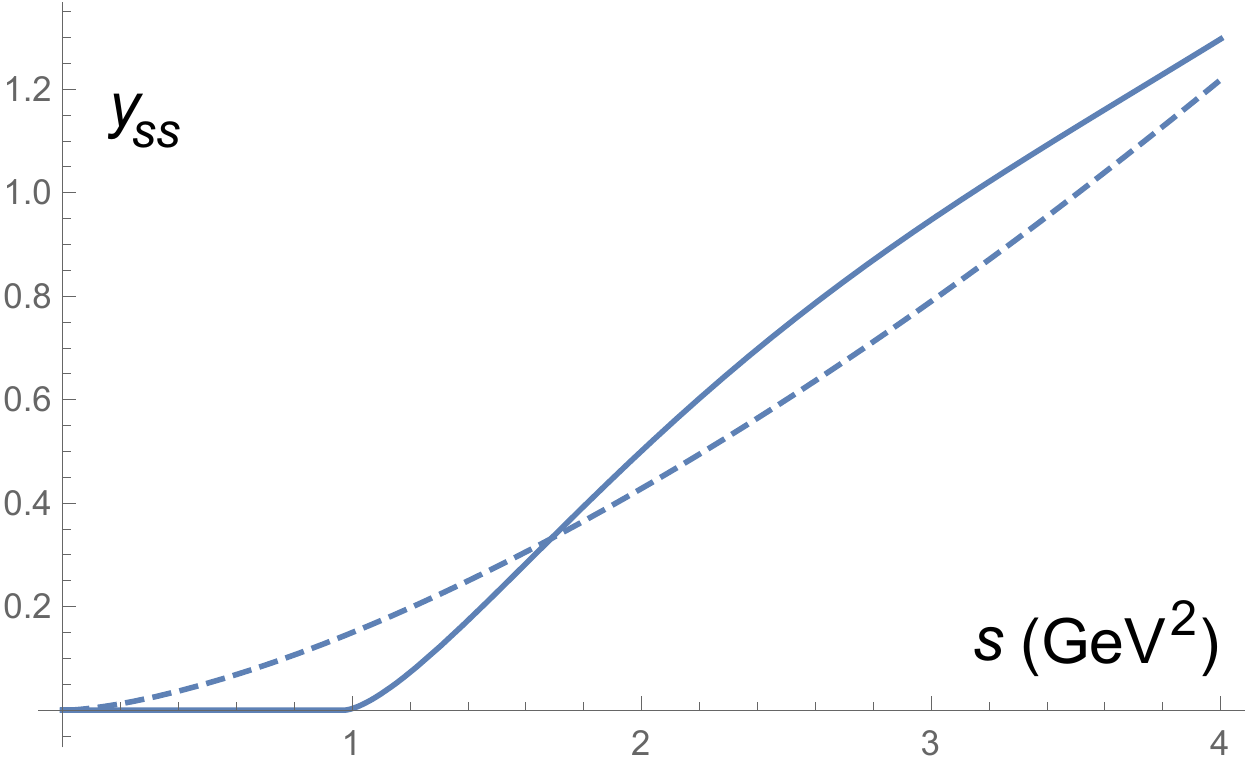}

(a) \hspace{5.0 cm} (b)\hspace{5.0 cm} (c)
\caption{\label{fig2}
Comparison of the solutions $y_{ij}(s)\equiv \Gamma_{ij}(s)/\Gamma$ (solid lines) with the inputs 
$y_{ij}^{\rm box}(s)\equiv \Gamma_{ij}^{\rm box}(s)/\Gamma$ (dashed lines) for 
(a) $ij=dd$, (b) $ij=ds$ and (c) $ij=ss$ at $\Lambda=5$ GeV$^2$.}
\end{center}
\end{figure}

Repeating the steps, we determine the solutions of the components $\Gamma_{dd}(s)$ and $\Gamma_{ss}(s)$ 
with the optimal choices $N=16$ and $N=10$, respectively, for the given $\Lambda=5$ GeV$^2$. The results 
of $\Gamma_{ij}(s)$ are compared with the inputs $\Gamma_{ij}^{\rm box}(s)$ in terms of their ratios over 
the total width $\Gamma$ in Fig.~\ref{fig2}. It is seen that the solutions maintain the monotonic increase 
of the input functions with $s$ basically, but the detailed behaviors have been modified by  
the physical thresholds. For $y_{dd}^{\rm (box)}(s)\equiv \Gamma_{dd}^{\rm (box)}(s)/\Gamma$, 
the hadron-level threshold $4m_\pi^2$ and the quark-level threshold $4m_d^2$ are both tiny, so the 
modification is minimal as shown in Fig.~\ref{fig2}(a). The thresholds $(m_\pi+m_K)^2$ and $4m_K^2$ for 
$y_{ds}$ and $y_{ss}\equiv \Gamma_{ss}(s)/\Gamma$ at the hadron level are not only much greater than 
$(m_d+m_s)^2$ and $4m_s^2$ at the quark level, respectively, but sizable. Therefore, the difference 
between the solutions and the inputs is more salient, as exhibited in Figs.~\ref{fig2}(b) and 
\ref{fig2}(c). The solutions stay vanishing till $s$ crosses the physical thresholds, such that their 
magnitudes at higher $s$ must be enhanced in order to compensate the loss at lower $s$, if the integral 
on the left-hand side of Eq.~(\ref{ij2}) remains equal to the right-hand side. This also explains why 
the enhancement is the most prominent in $y_{ss}(s)$, which is about 15\% around the $D$ meson mass 
squared $s=m_D^2\approx 3.5$ GeV$^2$, with the much larger threshold $4m_K^2\approx 1$ GeV$^2$. It is 
reasonable to claim 15\% violation of the quark-hadron duality in the channel with two strange quarks 
for the $D$ meson mixing, of the same order as postulated in \cite{Jubb:2016mvq}. All 
the solutions approach the inputs as $s\to\infty$, following the design in Eq.~(\ref{sub}). The 
aforementioned modifications originate from the nonperturbative effects characterized by the physical 
thresholds $m_{IJ}\not=m_{ij}$, whose introduction for the components $\Gamma_{ij}(s)$ in our formalism 
is unambiguous.

\begin{figure}
\begin{center}
\includegraphics[scale=0.4]{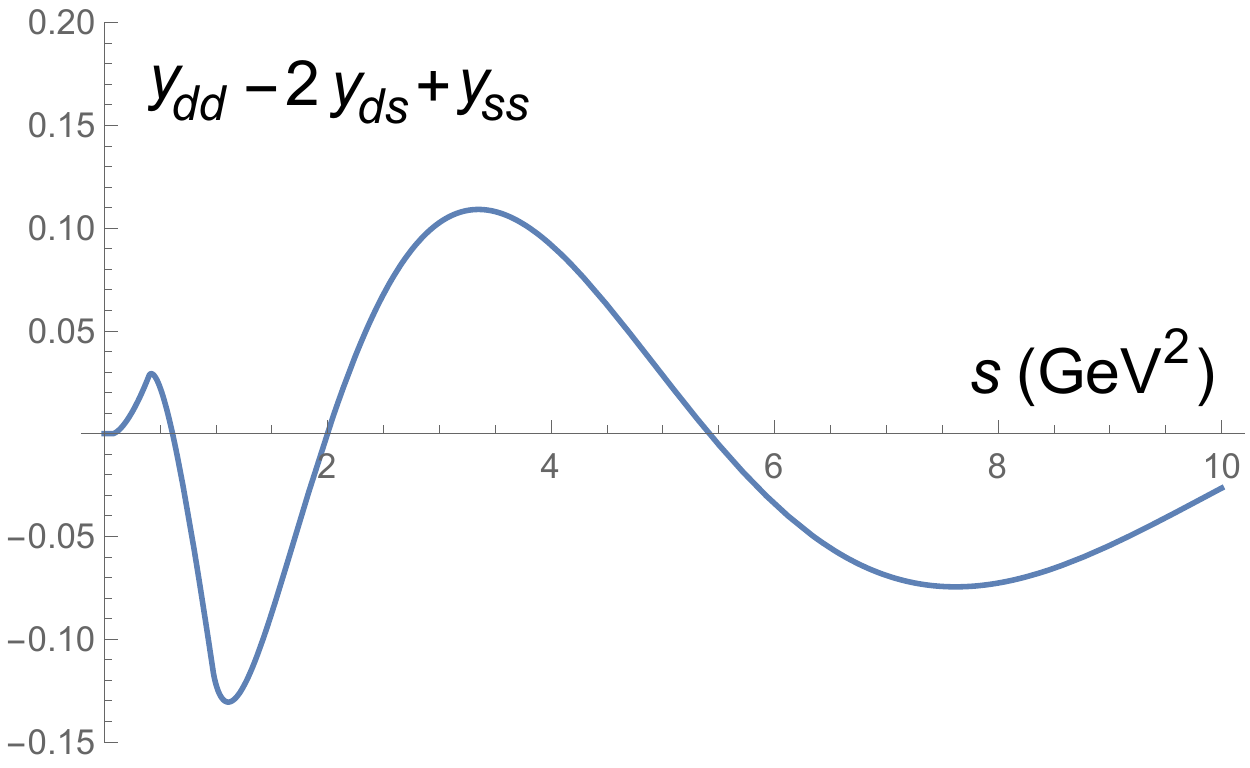}
\includegraphics[scale=0.4]{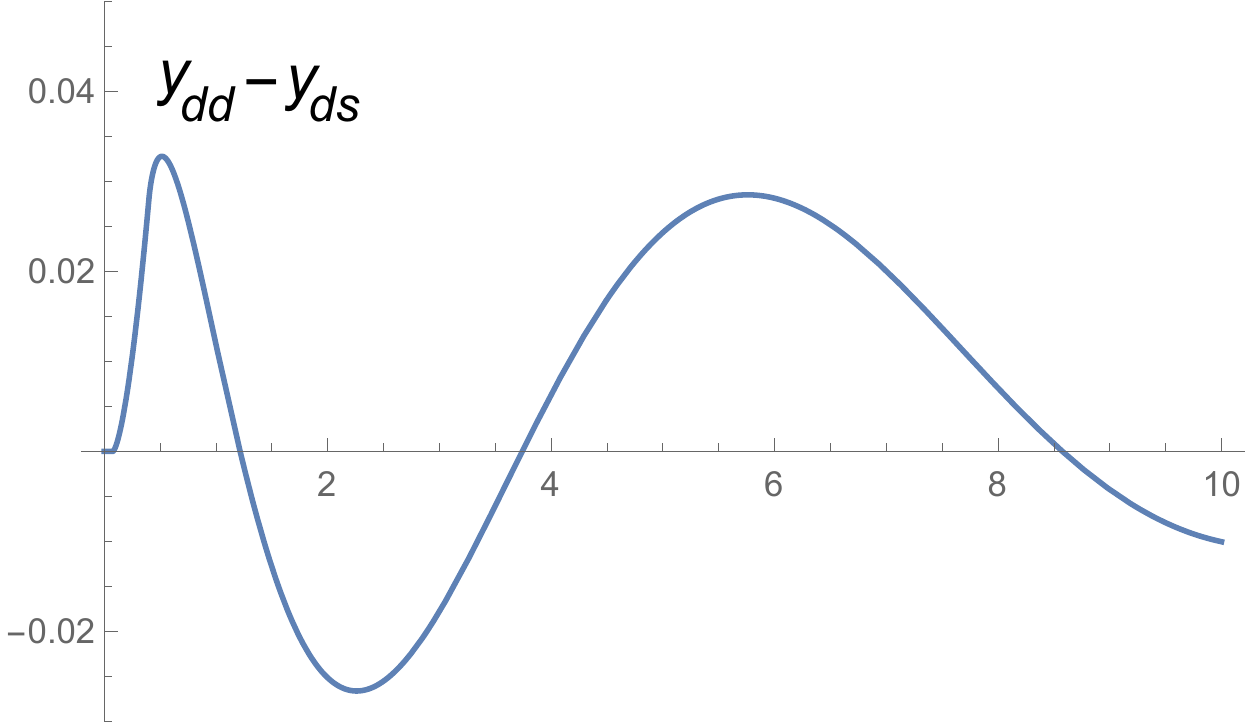}
\includegraphics[scale=0.4]{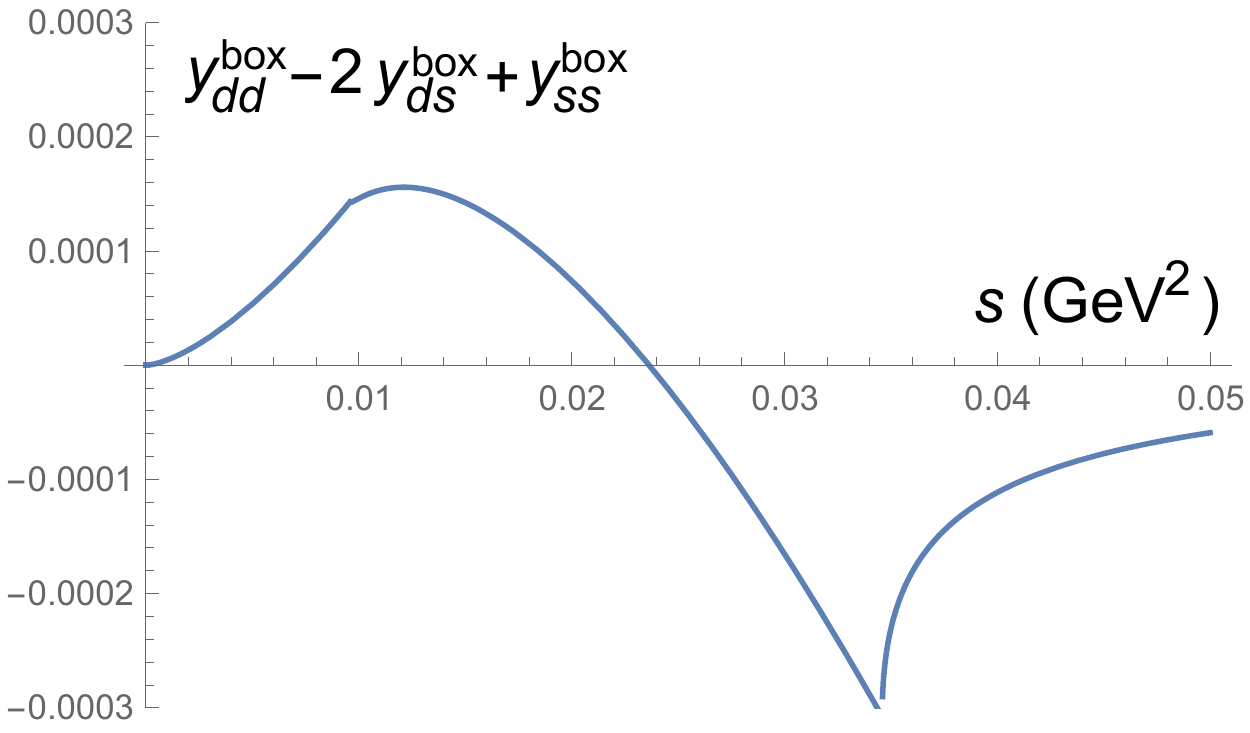}

(a) \hspace{5.0 cm} (b)\hspace{5.0 cm} (c)
\caption{\label{fig3}
Dependencies of (a) $y_{dd}-2y_{ds}+y_{ss}$, (b) $y_{dd}-y_{ds}$ and (c) 
$y_{dd}^{\rm box}-2y_{ds}^{\rm box}+y_{ss}^{\rm box}$ on $s$ for $\Lambda=5$ GeV$^2$.}
\end{center}
\end{figure}

We present in Fig.~\ref{fig3} the dependencies of the combinations $y_{dd}-2y_{ds}+y_{ss}$ and 
$y_{dd}-y_{ds}$ on $s$, which are associated with the CKM factors $\lambda_{s}^2$ and $\lambda_s\lambda_b$, 
respectively, for the given $\Lambda=5$ GeV$^2$. The oscillations of the 
curve in Fig.~\ref{fig3}(a) (not completely displayed in the plot) with the first peak (valley) located 
at $s\approx (m_\pi+m_K)^2$ ($s\approx 4m_K^2$) are anticipated \cite{Li:2020xrz}: when $s$ increases 
and crosses the threshold $(m_\pi+m_K)^2$ ($4m_K^2$), the single (double) strange quark channel with a 
destructive (constructive) contribution is opened, so the curve starts to descend (ascend). It is not 
difficult to understand the minor oscillations at higher $s$, since heavier states stemming from the $dd$, 
$ds$ and $ss$ channels are allowed to contribute in turn. These oscillations attenuate gradually, 
when the solutions for $\Gamma_{dd}(s)$, $\Gamma_{ds}(s)$ and $\Gamma_{ss}(s)$ approach the perturbative 
inputs at large $s$, as indicated in Fig.~\ref{fig2}, and the GIM suppression becomes effective. The curve 
for the combination $y_{dd}-y_{ds}$ in Fig.~\ref{fig3}(b) also reveals several oscillations but with 
smaller amplitudes, because of the stronger cancellation between $y_{dd}$ and $y_{ds}$ than between 
$y_{ds}$ and $y_{ss}$. This pattern can be interpreted by means of Fig.~\ref{fig2}, which shows the 
increasing enhancements from $y_{dd}$ to $y_{ds}$ and to $y_{ss}$ at $s= m_D^2$ compared with the inputs. 
Hence, the SU(3) symmetry breaking between the first two is smaller than between the last two. The first 
peak in Fig.~\ref{fig3}(b) appears at $s\approx (m_\pi+m_K)^2$ as expected, but the other peaks and 
valleys are shifted toward slightly higher $s$ compared to Fig.~\ref{fig3}(a): the constructive $ss$ 
channel is absent, so the descent of the curve cannot be reversed at $s\approx 4m_K^2$.

The aforementioned combinations of $y_{ij}$, where the box-diagram terms in Eq.~(\ref{sub}) cancel 
almost exactly, are in fact proportional to those of the subtracted functions $\Delta \Gamma_{ij}$. 
The results in Figs.~\ref{fig3}(a) and \ref{fig3}(b), multiplied by the corresponding 
CKM factors $\lambda_s^2$ and $2\lambda_s\lambda_b$, respectively, then behave differently from 
the box-diagram contributions: both pieces from the box diagrams are of $O(10^{-7})$ at 
the $D$ meson mass, but the $\lambda_s^2$ piece in our solutions becomes $O(10^{-3})$, and dominant over 
the $\lambda_s\lambda_b$ piece, which is of $O(10^{-6})$. The smallness of the latter is not only 
attributed to the shorter peak of $y_{dd}-y_{ds}$, but to its shift away from the $D$ meson mass, as 
shown in Fig.~\ref{fig3}(b). The $\lambda_b^2$ piece, depending on $y_{dd}$ in Fig.~\ref{fig2}(a),
is as tiny as $O(10^{-7})$. The SU(3) symmetry breaking effects from the various thresholds 
$m_{IJ}$, i.e., various enhancements in Figs.~\ref{fig2}(a)-\ref{fig2}(c), are manifested by
the dramatically different magnitudes of the combination $y_{dd}-2y_{ds}+y_{ss}$ in Fig.~\ref{fig3}(a)
and of $y_{dd}^{\rm box}-2y_{ds}^{\rm box}+y_{ss}^{\rm box}$ in Fig.~\ref{fig3}(c). Another feature 
of Fig.~\ref{fig3}(c) is that the shape of the curve is trivial: it reaches a peak at 
$s\approx (m_d+m_s)^2$ and a valley at $s\approx 4m_s^2$, and then approaches zero smoothly.

\begin{figure}
\begin{center}
\includegraphics[scale=0.47]{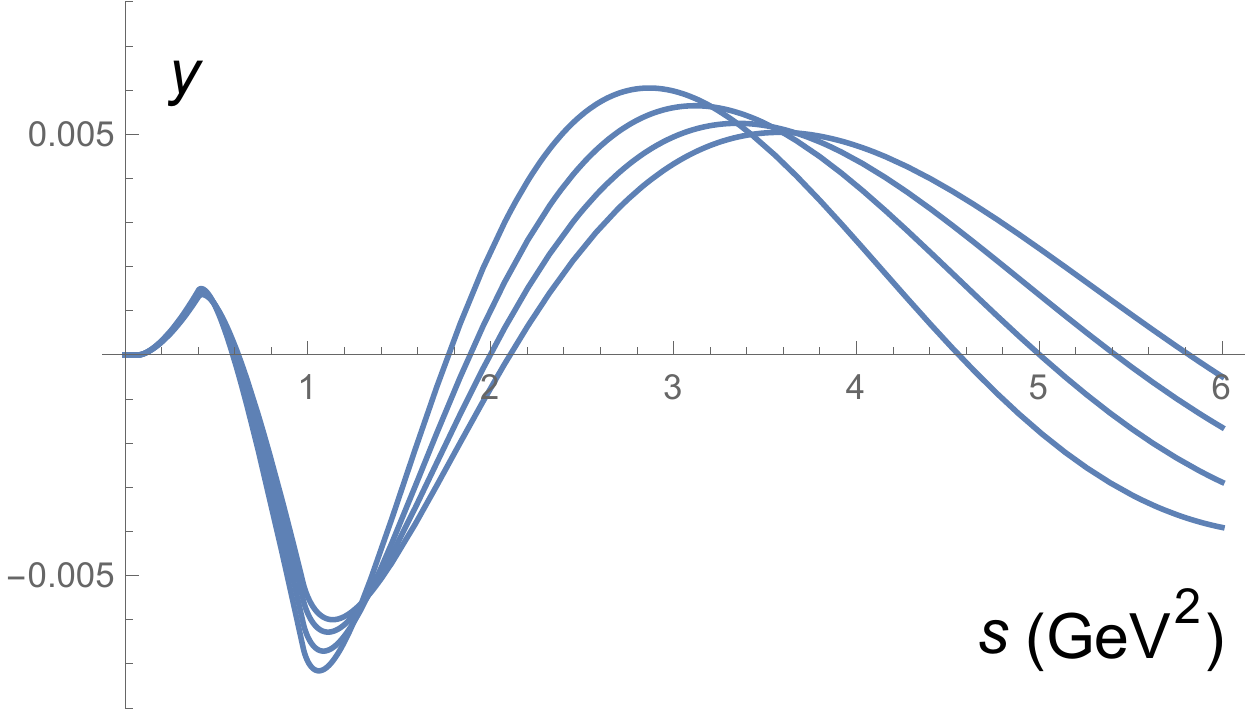}
\includegraphics[scale=0.47]{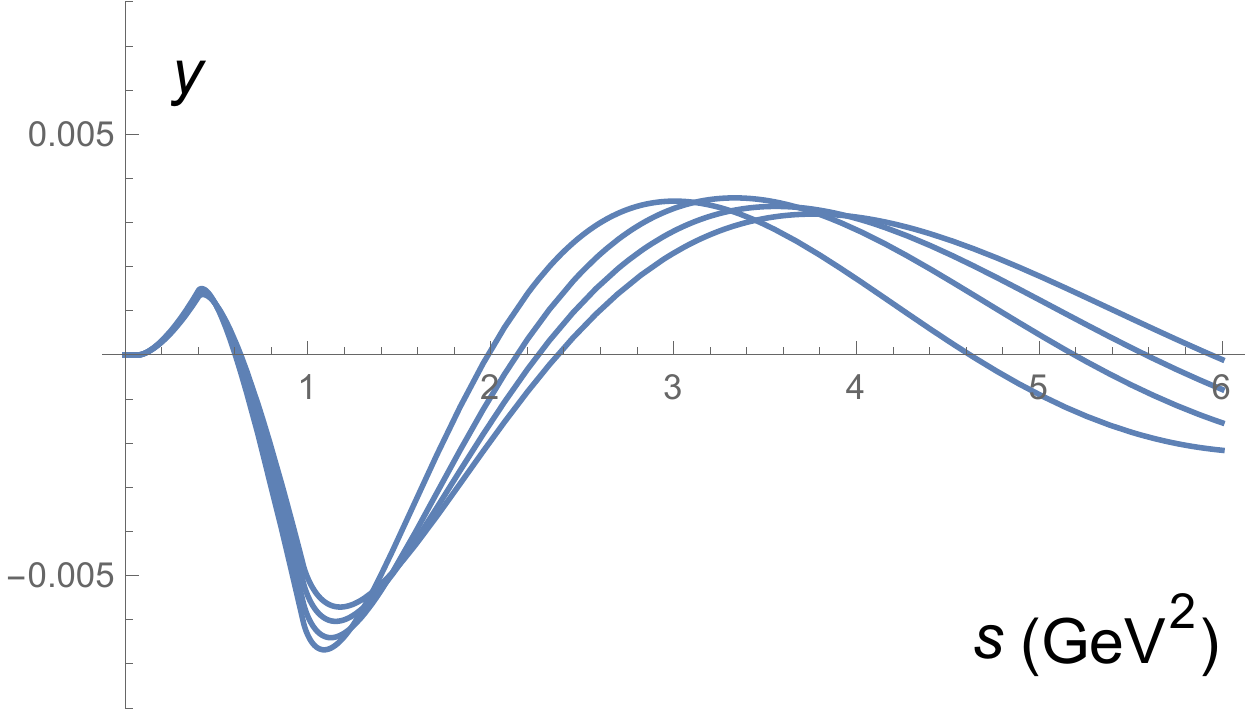}

(a) \hspace{5.0 cm} (b)
\caption{\label{fig4}
Solutions of $y(s)$ for $\Lambda=4.0$ GeV$^2$, 4.5 GeV$^2$, 5.0 GeV$^2$ and 5.5 GeV$^2$, corresponding to
the curves with the peaks from left to right, in the cases (a) with and (b) without the
second term in Eq.~(\ref{opei}).}
\end{center}
\end{figure}

We then investigate how the solutions for the mixing parameter $y(s)$ change with the transition scale 
$\Lambda$, starting from the CP-conserving case. The contributions from all the three pieces $\lambda_s^2$, 
$\lambda_s\lambda_b$ and $\lambda_b^2$ are included, though the behavior of $y(s)$ is governed by the 
first piece as stated above. It is encouraging to see in Fig.~\ref{fig4}(a) that the curves for 
$\Lambda=4.0$ GeV$^2$, 4.5 GeV$^2$, 5.0 GeV$^2$ and 5.5 GeV$^2$ all pass through the 
small region around $s\approx m_D^2$ and $y\approx 0.5\%$. Namely, a stability window 
in $\Lambda$ may exist, within which the obtained $y(m_D^2)$ is insensitive to $\Lambda$. The tails of 
these curves are far apart from each other, implying that they will not cross again at higher $s$. 
Therefore, $y(m_D^2)\approx 0.5\%$ is the unique solution from our method. Note that the above curves 
overlap completely in the region with $s < 1$ GeV$^2$, which, however, do not represent solutions for 
the physical $D$ meson apparently. To verify the postulation that the nonperturbative effects from the 
physical thresholds are crucial for stabilizing the solutions, we drop the second term in the 
input in Eq.~(\ref{opei}), and derive $y(s)$ for the same set of $\Lambda$ values in Fig.~\ref{fig4}(b). 
The curves have shapes similar to those in Fig.~\ref{fig4}(a), but scatter to some degree, 
such that the area in which they cross each other stretches. 
It means that the stability deteriorates in the absence of the nonperturbative effects. Besides, the 
magnitudes at $s\approx m_D^2$ reduce by about 40\%, which is the appropriate weight of nonperturbative
contributions to achieve the stability in QCD sum rules.




\begin{figure}
\begin{center}
\includegraphics[scale=0.5]{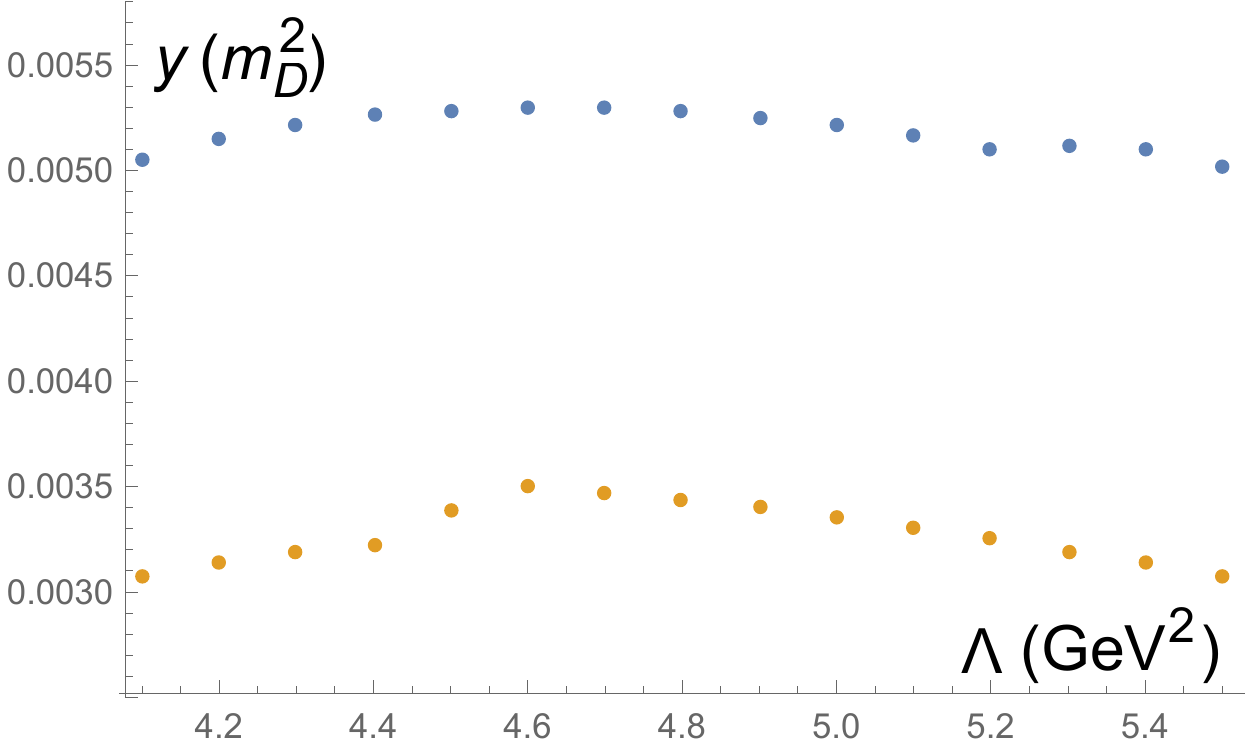}
\caption{\label{fig5}
Dependencies of $y(m_D^2)$ on $\Lambda$ in the cases with (upper curve) and without (lower curve) the
second term in Eq.~(\ref{opei}).}
\end{center}
\end{figure}

We read off the values of $y(m_D^2)$ at the $D$ meson mass squared $s=m_D^2$ from the curves like those in 
Fig.~\ref{fig4}, and plot the dependencies of $y(m_D^2)$ on the transition scale $\Lambda$ in the cases 
with and without the second term in Eq.~(\ref{opei}). It is noticed in
Fig.~\ref{fig5} that the former ascends with $\Lambda$ 
first, reaches a plateau around $\Lambda=4.5$ GeV$^2$, and then descends as $s>4.8$ GeV$^2$. Selecting the 
values in the range $\Lambda=[4.2,5.1]$ GeV$^2$ as our representative results, we have 
$y(m_D^2)=(0.52\pm 0.01)\%$, where the central value is located at $\Lambda=4.3$ GeV$^2$, and the tiny 
error reflects the remarkable stability of $y(m_D^2)$ with respect to the variation of $\Lambda$. For 
$s$ slightly below (above) $m_D^2$, say, $s=3.0$ GeV$^2$ ($s=4.0$ GeV$^2$), Fig.~\ref{fig4}(a) indicates 
that $y$ always decreases (increases) with $\Lambda$. The obtained $y(m_D^2)$, greater than in the 
exclusive analysis focusing only on two-body decays \cite{Jiang:2017zwr}, hints the sizable 
contributions from the resonances or multi-body states near the $D$ meson mass 
\cite{Golowich:1998pz,Falk:2001hx,Jiang:2017zwr}. As the nonperturbative effects are ignored, 
the plateau in $\Lambda$ disappears: the curve ascends with $\Lambda$, and then descends from the maximum 
located at $\Lambda=4.6$ GeV$^2$ directly, such that it is hard to extract any physical outcomes in this 
case. We stress that there is no free parameter in our approach, which can be tuned to achieve data 
fitting. The solutions of $y(m_D^2)$ are 
insensitive to the number $N$ for the polynomial expansion and to the arbitrary transition scale 
$\Lambda$ as stated before. We mention that the renormalization scales associated with the different 
channels for $y$ in the heavy quark expansion took different values so as to accommodate the data by 
avoiding the stringent GIM cancellation \cite{Lenz:2020efu}.



\begin{figure}
\begin{center}
\includegraphics[scale=0.5]{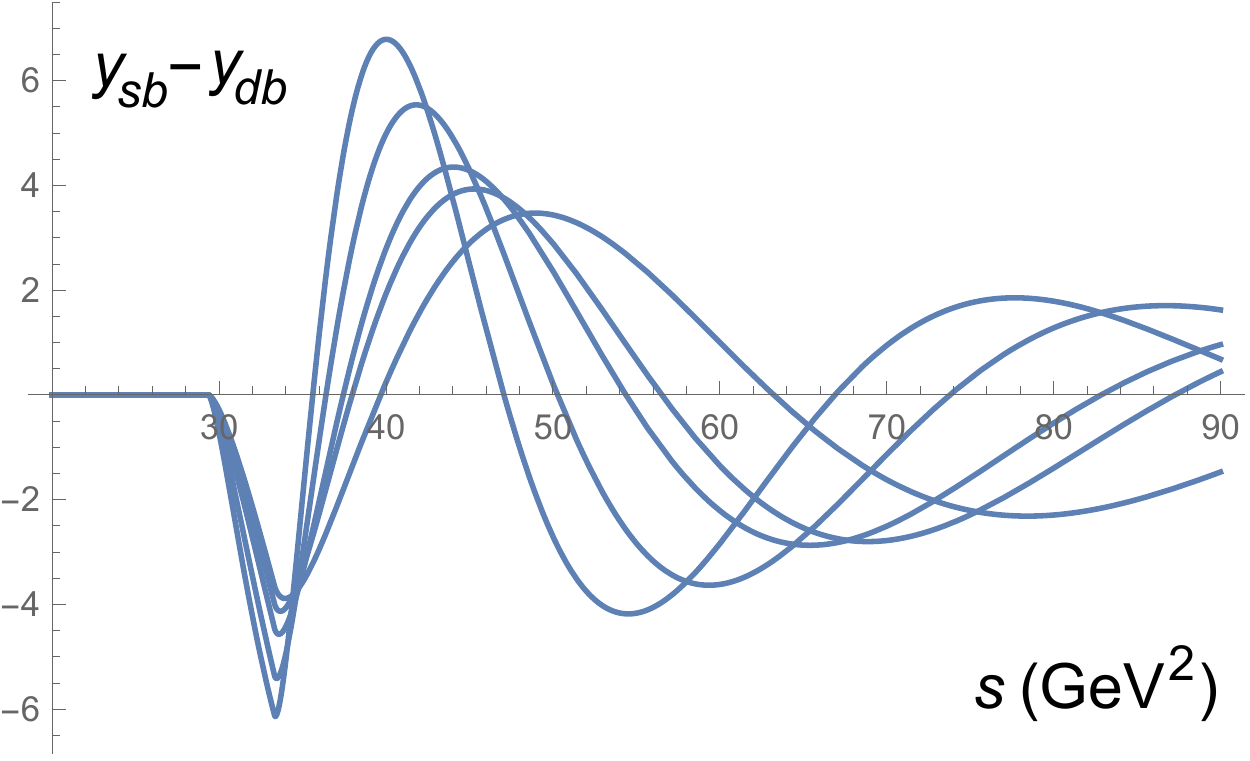}
\caption{\label{fig6} 
Behaviors of $y_{sb}-y_{db}\equiv(\Gamma_{sb}-\Gamma_{db})/\Gamma$ for $\Lambda=15$ GeV$^2$,
20 GeV$^2$, 25 GeV$^2$, 30 GeV$^2$ and 35 GeV$^2$, corresponding to the curves with the peaks
from left to right.}
\end{center}
\end{figure}

We then calculate the mixing parameter $x(m_D^2)$ according to Eq.~(\ref{xy}), for which the other three
components involving $b$ quarks, i.e., $\Gamma_{db}(s)$, $\Gamma_{sb}(s)$ and $\Gamma_{bb}(s)$ should be 
available first. Similarly, we seek the most convergent solutions in the polynomial 
expansion with the same power-law behaviors near the physical thresholds for the above three components. 
The dependences of the combination $y_{sb}-y_{db}\equiv(\Gamma_{sb}-\Gamma_{db})/\Gamma$ on $s$ for 
$\Lambda=15$ GeV$^2$, 20 GeV$^2$, 25 GeV$^2$, 30 GeV$^2$ and 35 GeV$^2$ are displayed 
in Fig.~\ref{fig6}. The curves run along the horizontal axis till the threshold near $m_B^2$ (the pion and 
kaon masses can be ignored here), and then oscillate, similar to the curves in Fig.~\ref{fig3}. It is seen 
that the curves corresponding to $\Lambda=25$ GeV$^2$ and 30 GeV$^2$ are relatively close to each other, 
revealing sort of stability. Since the result of $x(m_D^2)$ has little dependence on these components, 
which take substantial values at $s$ far away from $m_D^2$, we simply fix $\Lambda$ to 30 GeV$^2$, with
which $N=10$ ($N=11$) is chosen for $\Gamma_{db}(s)$ ($\Gamma_{sb}(s)$). The contribution to $x(m_D^2)$ 
from the component $\Gamma_{bb}(s)$ is even less important, so we also set $\Lambda=30$ GeV$^2$ for its 
evaluation for simplicity. It turns out that the contributions from the above three components to 
$x(m_D^2)$ via Eq.~(\ref{xy}) are as low as $2.2\times 10^{-7}$, among which $\Gamma_{bb}(s)$, 
contributing $O(10^{-10})$, is absolutely negligible.


\begin{figure}
\begin{center}
\includegraphics[scale=0.5]{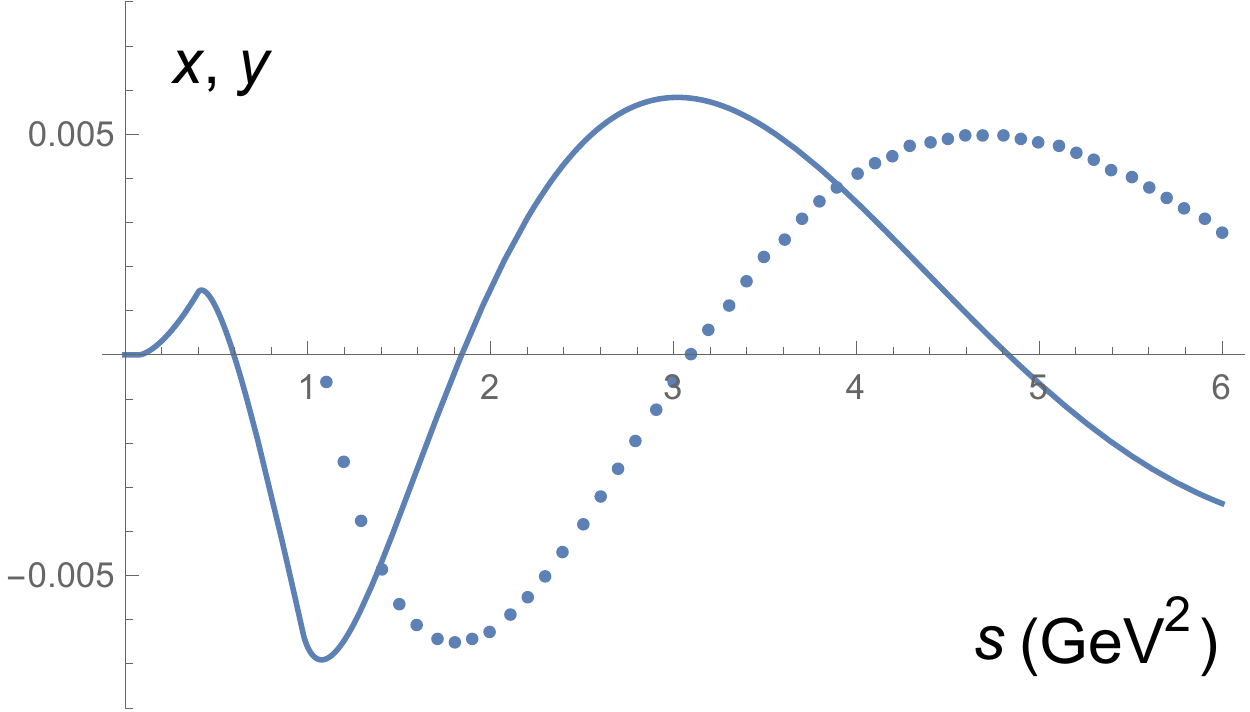}
\caption{\label{fig7} 
Behaviors of $x(s)$ (dotted line) and $y(s)$ (solid line) for $\Lambda=4.3$ GeV$^2$.}
\end{center}
\end{figure}

The curve of $y(s)$ for $\Lambda=4.3$ GeV$^2$, which gives rise to the central value of $y(m_D^2)$,
together with the corresponding $x(s)$ derived from Eq.~(\ref{xy}), are exhibited in Fig.~\ref{fig7}, 
from which we read off the central value $x(m_D^2)=0.21\%$. The correlation between $x(s)$ and 
$y(s)$ is similar to what was observed in \cite{Li:2020xrz}. The lower (upper) bound of $y(m_D^2)$ located 
at $\Lambda=4.2$ ($\Lambda=4.5$) GeV$^2$ leads to the upper (lower) bound of $x(m_D^2)=0.24\%$ ($0.15\%$). 
That is, we obtain $x(m_D^2)=(0.21^{+0.03}_{-0.06})\%$, whose error reflects the uncertainty in our method. 
We then survey the uncertainties from the theoretical inputs. The parameters involved in the CKM matrix and 
the hadron masses have been known precisely, so the associated uncertainties are minor. It has been
affirmed that our results are insensitive to the down quark mass $m_d$, and the $\pm 10\%$ variation of the
strange quark mass $m_s$ induces only $\mp 0.6\%$ error to the value of $y(m_D^2)$. The uncertainty form 
the overall hadronic parameters, like the bag parameters, is about 5\% according to
\cite{Carrasco:2014uya,Carrasco:2015pra}.
We present our predictions for the mixing parameters in the $CP$-conserving case, including the overall 
5\% uncertainty, as
\begin{eqnarray}
x(m_D^2)=(0.21^{+0.04}_{-0.07})\%,\;\;\;\;y(m_D^2)=(0.52\pm 0.03)\%.\label{cpi}
\end{eqnarray}
It is emphasized that the uncertainties from 
neglected subleading contributions to the inputs at large mass 
have not been taken into account. According to \cite{Lenz:2006hd}, the $O(\alpha_s)$ and 
$O(1/m_b)$ corrections, $\alpha_s$ being the strong coupling constant, amount to about 20\% of the 
leading contribution to the $B_s$ meson width difference. It is thus likely that the results
in Eq.~(\ref{cpi}) suffer additional uncertainties of order 20\%.


It has been shown in lattice analyses \cite{Bazavov:2017weg} that the $(S-P)(S-P)$ hadronic matrix element 
for the $D$ meson mixing is larger than the $(V-A)(V-A)$ one. This observation is understandable, because 
the former is proportional to an additional factor $m_D^2/(m_c+m_u)^2\approx 2$ actually, $m_u$ being the 
$u$ quark mass. Therefore, it is possible to gain an overall 30\% enhancement of our predictions by 
considering the above factor, which then agree with the data \cite{HFLAV:2022pwe}
\begin{eqnarray}
x=(0.44^{+0.13}_{-0.15})\%,\;\;\;\;
y=(0.63\pm 0.07)\%,\label{data}
\end{eqnarray}
in the $CP$-conserving case. Our goal is not to achieve an exact fit to the data, but to
demonstrate that the box-diagram contribution to the $D$ meson mixing can be amplified by a factor of
$10^4$ under the nonperturbative SU(3) breaking effects. A precise study can be carried out by employing
the weak effective Hamiltonian and the hadronic matrix elements with higher accuracy
for the perturbative inputs in our formalism.




The $CP$ violation in the $D$ meson mixing has been discussed and formulated in detail in 
\cite{Kagan:2020vri}. In the $CP$ violating case we simply multiply our solution for each component 
$\Gamma_{ij}$ in Eq.~(\ref{dec}) by the associated complex CKM factors, and both $M_{12}$ and 
$\Gamma_{12}$ become complex. We then adopt the general definitions of 
the mixing parameters $x$ and $y$ in Eq.~(\ref{3}), and find that the imaginary parts of $M_{12}$ and 
$\Gamma_{12}$, being of $Q(10^{-3})$ and $O(10^{-4})$ of the real parts, respectively, are negligible. 
Hence, our predictions for $x$ and $y$ remain the same as in Eq.~(\ref{cpi}) basically, and will be 
close to the data
\begin{eqnarray}
x=(0.409^{+0.048}_{-0.049})\%,\;\;\;\; y=(0.615^{+0.056}_{-0.055})\%,
\end{eqnarray} 
in the $CP$-violating case \cite{HFLAV:2022pwe}, after the enhancement from the $(S-P)(S-P)$ hadronic 
matrix element is taken into account. We also derive 
\begin{eqnarray}
\left|\frac{q}{p}\right|-1= (-3.0^{+0.1}_{-0.0})\times 10^{-4},\;\;\;\;
Arg\left(\frac{q}{p}\right)= (3.1^{-0.3}_{+0.4})^\circ\times 10^{-3},
\end{eqnarray}
where the central values (the upper errors, the lower errors) come from the scales 
$\Lambda=4.3$ GeV$^2$ ($\Lambda=4.2$ GeV$^2$,  $\Lambda=4.5$ GeV$^2$). They can be compared with 
the measured values $|q/p|=0.995 \pm 0.016$  and $Arg(q/p) = (-2.5\pm 1.2)^\circ$ 
\cite{HFLAV:2022pwe}, which were obtained under the same phase convention for the CP transformation 
of neutral $D$ mesons, and help constrain new physics models \cite{Ball:2007yz,Xing:2007sd} due to
their small theoretical uncertainties. Besides, we predict the quantity 
$\phi_{12}\equiv Arg(M_{12}/\Gamma_{12})\approx -0.049^\circ$ in accordance with
the data $\phi_{12}=(0.58^{+0.91}_{-0.90})^\circ$ \cite{HFLAV:2022pwe}.


\section{$B_{s(d)}$ MESON MIXING AND KAON MIXING}

Tremendous efforts have been devoted to perturbative studies of the $B_{d(s)}$ meson mixing and the kaon 
mixing, and to their confrontation with data in the literature. The transition matrix 
elements $M_{12}^{s(d)}-i\Gamma_{12}^{s(d)}/2$ for the $B_{d(s)}$ meson mixing have been 
evaluated up to two-loop QCD corrections in the heavy quark expansion 
\cite{Buras:1990fn,Gerlach:2022wgb}. The ratio of the width difference over the mass difference,
$\Delta\Gamma_{s(d)}/\Delta M_{s(d)}=Re(\Gamma_{12}^{s(d)}/M_{12}^{s(d)})$, where hadronic uncertainties 
largely cancel, was computed in \cite{Gerlach:2022wgb}. The experimental input 
$\Delta M_{s(d)}^{\rm exp}$ was then inserted to predict $\Delta\Gamma_{s(d)}$, which was shown to be 
consistent with the data. It implies that the $B_{s(d)}$ meson mixing can be accommodated by short-distance 
dynamics within hadronic uncertainties. A similar conclusion on the dominance of short-distance dynamics in 
the measured kaon mass difference was also drawn \cite{Herrlich:1996vf,Buras:2010pza,Brod:2011ty}. 
Therefore, we do not attempt precise explanations of the $B_{s(d)}$ meson mixing and the kaon mixing in 
this paper, on which a lot of progresses have been made, but corroborate that the neutral meson mixing, 
no matter whether it is governed by perturbative or nonperturbative dynamics, can be addressed consistently 
and systematically in our framework.


We decompose the absorptive piece of the transition matrix elements for the $B_{s(d)}$ meson mixing into
\begin{eqnarray}
\Gamma_{12}^{s(d)}(m_{B_{s(d)}}^2)=\lambda_{u}^{s(d)2}\Gamma_{uu}^{s(d)}(m_{B_{s(d)}}^2)
+2\lambda_u^{s(d)}\lambda_c^{s(d)}\Gamma_{uc}^{s(d)}(m_{B_{s(d)}}^2)
+\lambda_{c}^{s(d)2}\Gamma_{cc}^{s(d)}(m_{B_{s(d)}}^2),\label{s12}
\end{eqnarray}
to which a top quark does not contribute, with the CKM factors $\lambda_{u}^{s(d)}=V_{ub}V^*_{us(d)}$ and 
$\lambda_c^{s(d)}=V_{cb}V^*_{cs(d)}$, and the meson mass $m_{B_d}$. The box-diagram contributions 
$\Gamma_{ij}^{s(d)\rm box}(s)$ are the same as Eq.~(\ref{CKM}), but with the replacements of $f_D$ 
($m_D$, $B_D$) by  $f_{B_{d(s)}}$ ($m_{B_{d(s)}}$, $B_{B_{d(s)}}$), and of $m_d$ ($m_s$) by  
$m_u$ ($m_c$). They yield the width 
difference $\Delta\Gamma_{s}=0.099$ ps$^{-1}$ from Eq.~(\ref{s12}) for the $B_s$ meson mass (decay 
constant) $m_{B_s}=5.367$ GeV ($f_{B_s}=0.230$ GeV) \cite{PDG}, the quark masses $m_u=0.005$ GeV and 
$m_c=1.3$ GeV, and the typical bag parameter $B_{B_s}=1$, close to the observed value 
$\Delta\Gamma_s^{\rm exp}=(0.084\pm 0.005)$ ps$^{-1}$ \cite{HFLAV:2022pwe} as stated above. The components 
$\Gamma_{uu}^{s(d)}(s)$, $\Gamma_{uc}^{s(d)}(s)$ and $\Gamma_{cc}^{s(d)}(s)$ for a fictitious 
$B_{s(d)}$ meson with the invariant mass squared $s$ will be derived in our method. The solutions for
the $B_s$ and $B_d$ mesons are expected to be very similar, so the distinction between 
$\Gamma_{12}^{s}(s)$ and $\Gamma_{12}^{d}(s)$ mainly comes from the CKM factors. Because of the 
hierarchy $|\lambda_{u}^s|\ll \lambda_c^s$, no cancellation occurs among the three pieces in 
Eq.~(\ref{s12}), such that the last term dominates $\Gamma_{12}^s(m_{B_s}^2)$, as having been noticed in 
\cite{Lenz:2011zz}. As to the $B_d$ meson mixing, for which $|\lambda_{u}^d|$ and $|\lambda_c^d|$ 
are of the same order of magnitude but unequal, a milder cancellation exists, and all the 
three pieces in Eq.~(\ref{s12}) contribute to $\Gamma_{12}^d(m_{B_d}^2)$. 


\begin{figure}
\begin{center}
\includegraphics[scale=0.4]{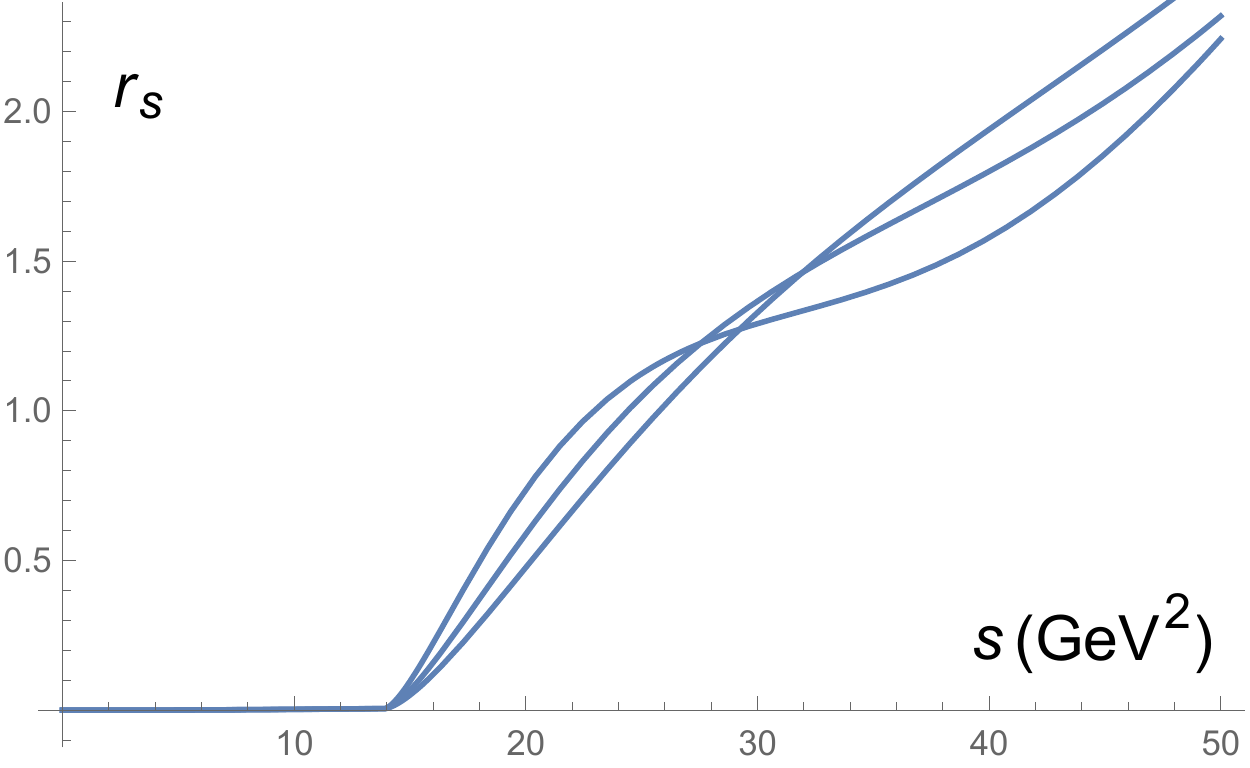}
\includegraphics[scale=0.4]{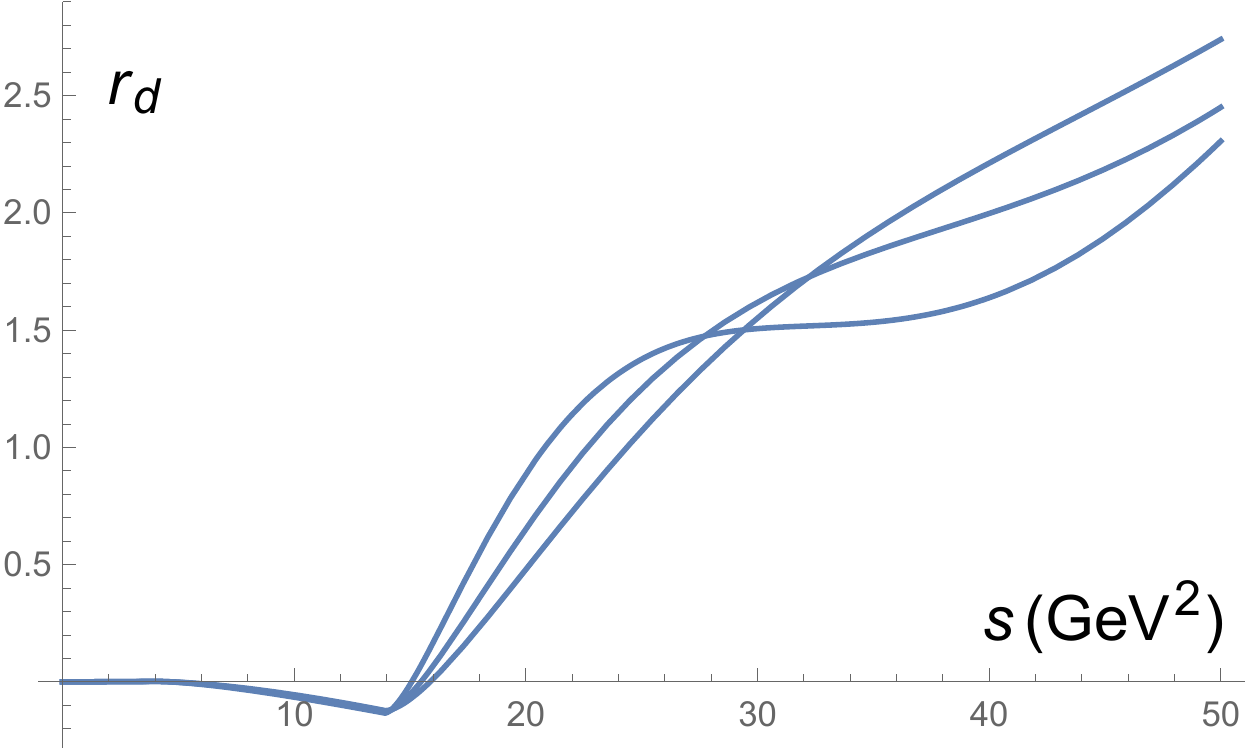}
\includegraphics[scale=0.4]{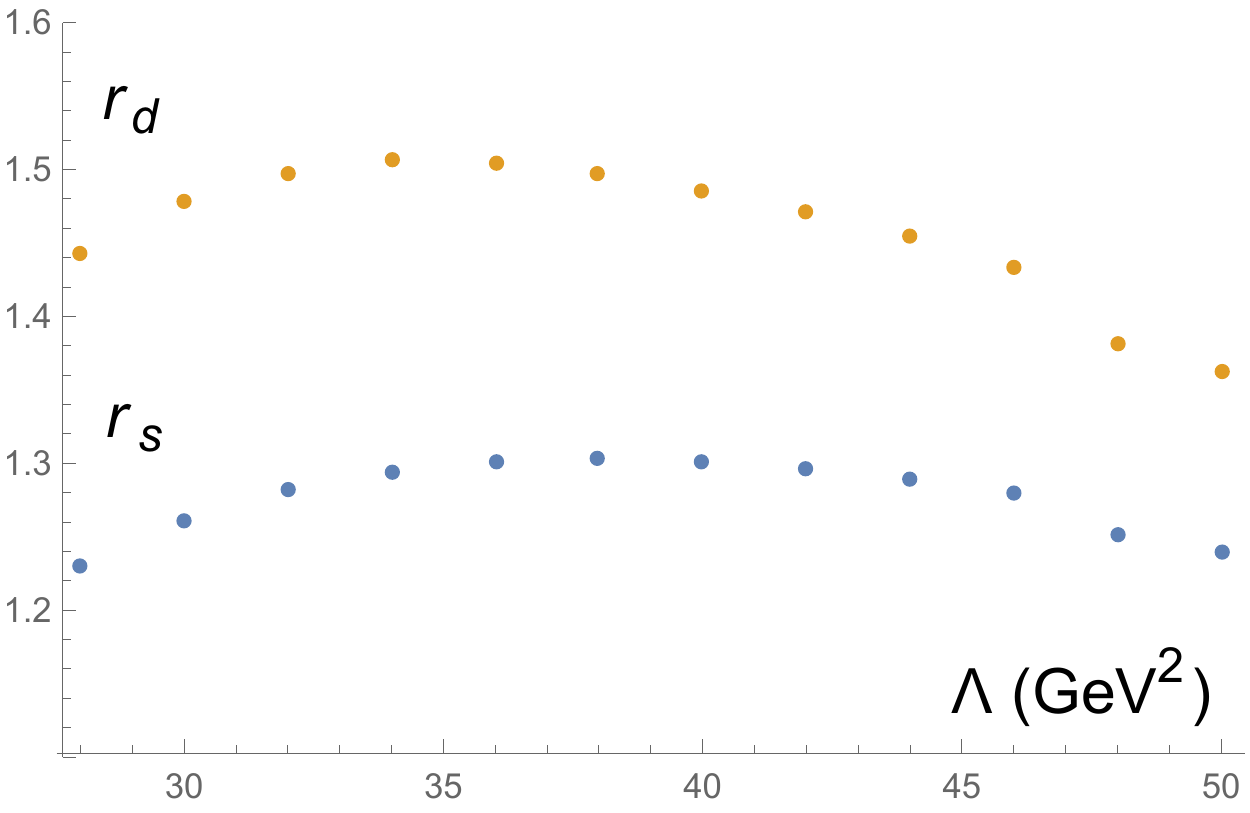}

(a) \hspace{5.0 cm} (b)\hspace{5.0 cm} (c)
\caption{\label{fig8}
Solutions of (a) $r_s(s)$ and (b) $r_d(s)$ for $\Lambda=30$ GeV$^2$, 40 GeV$^2$ and 50 GeV$^2$, 
corresponding to the curves which rise at the threshold $s=4m_D^2$ from left to right. (c)
Dependencies of $r_{s}(m_{B_{s}}^2)$ and $r_d(m_{B_d}^2)$ on $\Lambda$.}
\end{center}
\end{figure}

The construction of the dispersion relation for the $B_{d(s)}$ meson mixing follows the steps in Sec.~II, 
and the aforementioned $\Gamma_{ij}^{s(d)\rm box}(s)$ from the box diagrams are taken as the inputs.
We solve for the unknown vector $a^{(ij)}$ with the input vector $b^{(ij)}$ in 
Eq.~(\ref{opei}) to get the components $\Gamma_{uu}^{s(d)}(s)$, $\Gamma_{uc}^{s(d)}(s)$ and 
$\Gamma_{cc}^{s(d)}(s)$ for various transition scales $\Lambda$ as in the previous section. Here we 
consider the ratio
\begin{eqnarray}
r_{s(d)}(s)\equiv\frac{\Gamma_{12}^{s(d)}(s)}{\Gamma^{s(d)\rm box}_{12}(m_{B_{s(d)}}^2)},\label{rds}
\end{eqnarray} 
which is free of the hadronic uncertainties from the decay constant $f_{B_{d(s)}}$ and the bag 
parameters $B_{B_{d(s)}}$. It will be seen that there are stable solutions of order of unity for the 
ratio $r_{s(d)}(m_{B_{s(d)}}^2)$. In other words, the obtained $B_{s(d)}$ meson width difference 
does not deviate from the box-diagram contribution much under the nonperturbative effects. 
We mention that the quark-hadron duality, i.e., the equivalence between the quark-level and hadron-level 
evaluations of the $B_s$ meson width difference, has been demonstrated in \cite{Chua:2011er}.


We focus only on the $CP$-conserving case by picking up the real part of the CKM matrix element $V_{ub}$, 
and adopt the quark masses $m_u=0.005$ GeV and $m_c=1.3$ GeV, and the meson masses $m_{B_s}=5.369$ GeV and 
$m_{B_d}=5.280$ GeV \cite{PDG}. The best convergence of the polynomial expansion associated with the 
component $\Gamma_{uu}^{s(d)}(s)$ ($\Gamma_{uc}^{s(d)}(s)$, $\Gamma_{cc}^{s(d)}(s)$ fixes the optimal 
numbers $N=16$, 16 and 16 ($N=14$, 15 and 15, $N=11$, 11 and 10) for $\Lambda=30$ GeV$^2$, 40 GeV$^2$ 
and 50 GeV$^2$, respectively. It is encouraging to find that the three curves of $r_{s}$ ($r_{d}$) for 
the above $\Lambda$ values cross each other in the small region around $s\approx m_{B_{s(d)}}^2$ and 
$r_s\approx 1.3$ ($r_d\approx 1.5$) in Fig.~\ref{fig8}(a) (Fig.~\ref{fig8}(b)). Namely, a stability 
window in $\Lambda$ is present, within which the solutions of $r_{s(d)}(m_{B_{s(d)}}^2)$ are insensitive 
to $\Lambda$. This feature is similar to that of $y(m_D^2)$ in Fig.~\ref{fig5}. Since all the three 
pieces on the right-hand side of Eq.~(\ref{s12}) contribute to $r_{d}(s)$, it does not vanish below 
the threshold $s=4m_D^2$ as displayed in Fig.~\ref{fig8}(b). Moreover, the width difference for the 
$B_d$ meson is about $\lambda^2\sim 0.05$ of the $B_s$ meson one in agreement with the data 
\cite{HFLAV:2022pwe}.

We read off the values of $r_{s(d)}(m_{B_{s(d)}}^2)$, and plot 
its dependence on the scale $\Lambda$ in Fig.~\ref{fig8}(c). It is noticed that the curve for 
$r_{s(d)}(m_{B_{s(d)}}^2)$ ascends with $\Lambda$ first, becomes relatively flat around $\Lambda=38$ 
GeV$^2$ ($\Lambda=34$ GeV$^2$), where the maximum is located, and then descends monotonically. Selecting 
the values in the intervals $\Lambda=[32,46]$ GeV$^2$ and $\Lambda=[30,42]$ GeV$^2$ as our representative 
results, we have
\begin{eqnarray}
r_{s}(m_{B_s}^2)=1.29\pm 0.01,\;\;\;\;
r_{d}(m_{B_d}^2)=1.49\pm 0.02, \label{rsd}
\end{eqnarray}
respectively, whose tiny errors reflect the excellent stability of our solutions in the wide ranges of 
$\Lambda$. The solution for $\Gamma_{12}^{s}(m_{B_{s}}^2)$ is indeed of the same order as the input 
$\Gamma_{12}^{s\rm box}(m_{B_{s}}^2)$, as indicated by Eq.~(\ref{rsd}). The value $r_{d}(m_{B_d}^2)$ 
is slightly larger owing to the partial cancellation among the perturbative contributions to the three 
pieces in Eq.~(\ref{s12}). The above investigation confirms that the nonperturbative effects associated 
with the physical thresholds do not impact much the width differences, the quark-hadron duality holds 
reasonably well for the $B_{s(d)}$ meson mixing, and short-distance dynamics dominates the relevant 
observables.

\begin{figure}
\begin{center}
\includegraphics[scale=0.47]{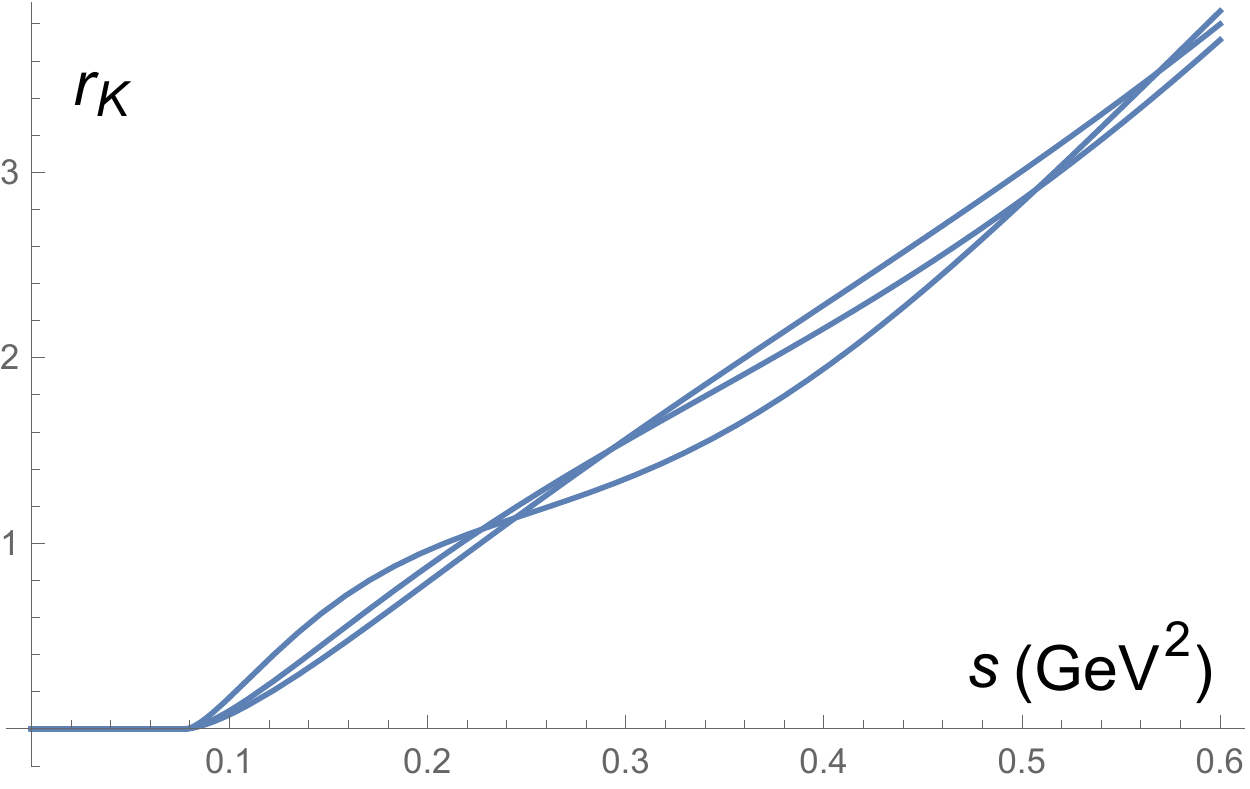}
\includegraphics[scale=0.47]{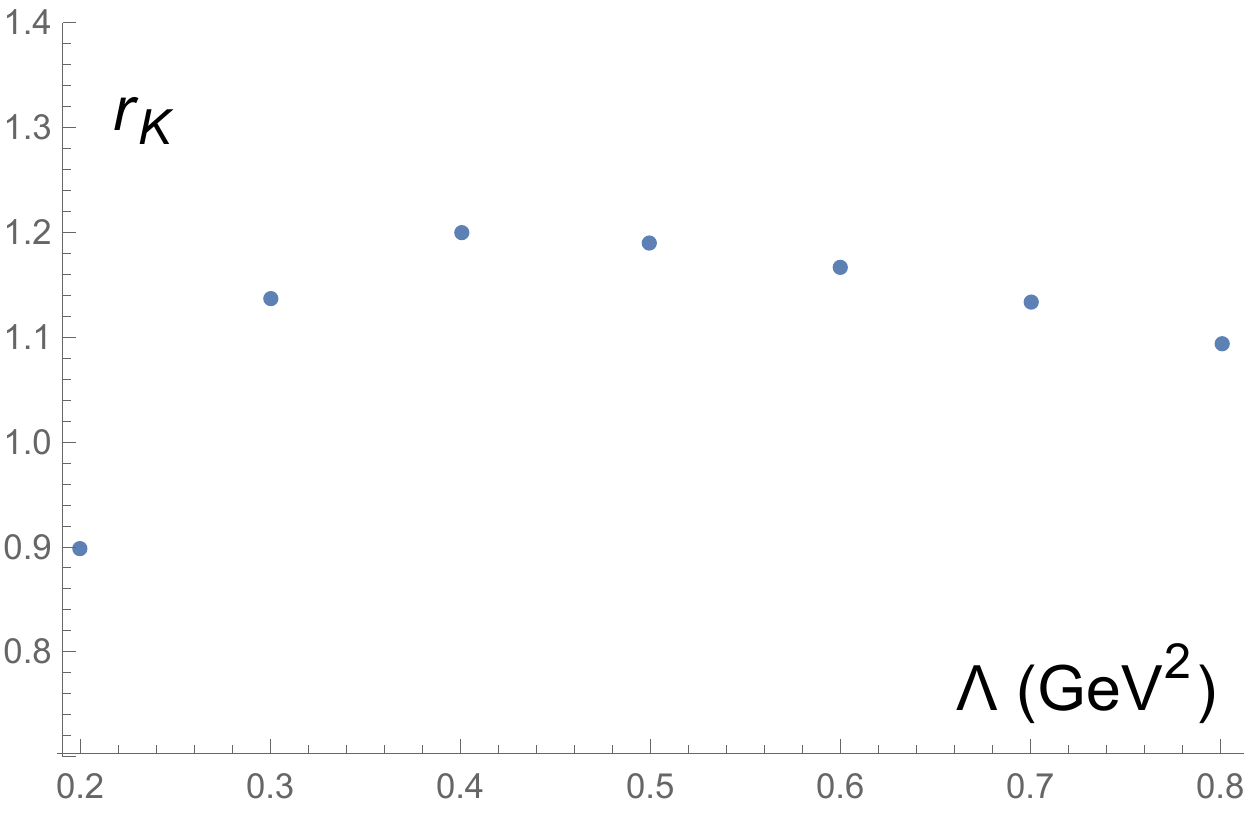}

(a) \hspace{5.0 cm} (b)
\caption{\label{fig9}
(a) Solutions of $r_K(s)$ for $\Lambda=0.3$ GeV$^2$, 0.5 GeV$^2$ and 0.7 GeV$^2$, corresponding to
the curves which rise from the threshold $s=4m_\pi^2$ from left to right. (b)
Dependence of $r_{K}(m_K^2)$ on $\Lambda$.}
\end{center}
\end{figure}

The absorptive piece $\Gamma_{12}^K(s)$ for a fictitious kaon with the invariant mass squared $s$ is also 
decomposed in terms of the CKM factors $\lambda_k\equiv V_{ks}V^*_{kd}$, $k=u,c,t$. Only the component 
$\Gamma_{uu}^K(m_K^2)$ contributes to $\Gamma_{12}^K(m_K^2)$, because a kaon does not decay into final 
states with charm quarks. The expression of the corresponding input $\Gamma_{uu}^{K\rm box}(s)$ at large 
$s$ is the same as Eq.~(\ref{CKM}), but with the appropriate replacements of the decay constant, the 
particle masses, and the bag parameter. Note that the $CP$ transformation sets 
$CP|D^0\rangle=-|\bar D^0\rangle$ in Sec.~II , so $K_1$ ($K_2$) refers to $K_L$ ($K_S$) in the analogous
convention. The width difference is almost equal to the width of $K_S$, i.e, 
$\Gamma_2^K-\Gamma_1^K=2\Gamma_{12}^K\approx \Gamma_2^K=7.347\times 10^{-15}$ 
GeV in experiments \cite{PDG}. It is straightforward to check that the box-diagram contribution is lower,
but accounts for the measured width difference at the order of magnitude. The higher-order QCD corrections
to the effective weak Hamiltonian \cite{SVZ77} and the penguin contribution \cite{D84} can be included into 
the input for a more precise analysis. Here we simply calculate the ratio 
\begin{eqnarray}
r_{K}(s)\equiv\frac{\Gamma_{12}^{K}(s)}{\Gamma_{12}^{K\rm box}(m_K^2)},\label{rk}
\end{eqnarray} 
with the denominator being derived from the box diagrams, and examine whether the ratio is of order of 
unity. 


We solve for the component $\Gamma_{uu}^K(s)$ for various scales $\Lambda$ by repeating the procedures,
and search for stable solutions of $r_K(m_K^2)$. The best convergence of the polynomial expansion 
associated with $\Gamma_{uu}^K(s)$ determines the numbers $N=11$, 10 and 12 for $\Lambda=0.3$ GeV$^2$, 
0.5 GeV$^2$ and 0.7 GeV$^2$, respectively. The corresponding results of $r_K(s)$ are exhibited in 
Fig.~\ref{fig9}(a), whose curves cross each other in the small region around $s\approx m_K^2$ and 
$r_K\approx 1.2$. That is, a stability window in $\Lambda$ can be identified, within which $r_K(m_K^2)$ 
is insensitive to $\Lambda$. We acquire the solutions for $r_K(s)$, read off the values of 
$r_K(m_K^2)$, and present its dependence on $\Lambda$ in Fig.~\ref{fig9}(b). The curve, with the shape 
similar to that of $r_{s(d)}(m_{B_{s(d)}}^2)$ in Fig.~\ref{fig8}(c), is relatively flat 
around $\Lambda=0.4$ GeV$^2$. Selecting $r_K(m_K^2)$ in the interval $\Lambda=[0.3,0.7]$ GeV$^2$ as our 
representative results, we get
\begin{eqnarray}
r_K(m_K^2)=1.17\pm 0.03,\label{rkv}
\end{eqnarray}
which, close to unity, hints the importance of short-distance contributions 
in the kaon mixing. We point out that the scale $\Lambda$ also 
bears the meaning of a resolution power of the inverse matrix method \cite{Li:2021gsx}, so
$\Lambda$ takes values near the resonance to be explored. It is then realized why the stability 
window appears at $\Lambda$ about hundreds of MeV$^2$, few GeV$^2$ and tens of GeV$^2$ in the kaon mixing, 
the $D$ meson mixing and the $B_{s(d)}$ meson mixing, respectively. 


At last, we summarize our observations on the neutral meson mixing mechanism, and highlight the uniqueness
of the charm mixing. For a more transparent illustration, we reexpress the absorptive piece of the 
transition matrix elements for the charm mixing as
\begin{eqnarray}
\Gamma_{12}(m_D^2)=\lambda_{d}^{2}\Gamma_{dd}(m_D^2)
+2\lambda_d\lambda_s\Gamma_{ds}(m_D^2)
+\lambda_{s}^2\Gamma_{ss}(m_D^2),\label{c12}
\end{eqnarray}
which turns the cancellation among the decay channels in Eq.~(\ref{g12}) into the cancellation among the 
components associated with the three CKM factors, because of the relation 
$\lambda_{d}^{2}\approx-\lambda_d\lambda_s\approx\lambda_{s}^2$. In fact, at most 15\% duality violation 
for each component in the $D$ meson mixing is less severe than in the others given in Eqs.~(\ref{rsd}) 
and (\ref{rkv}). Comparing $2m_K$ with $m_D$ and $2m_D$ with $m_{B_s}$, we see that the physical 
threshold is further below the neutral meson mass in the $D$ meson mixing than in the $B_s$ meson mixing. 
It is the reason why the deviation of the solution from the perturbative input caused by the threshold 
is minor in the former. As shown in \cite{Bobrowski:2012jf}, the charm width difference receives 
corrections from next-to-leading order QCD below 50\%, and $1/m_c$ corrections of 30\%. That is, 
subleading contributions in the charm mixing do not reveal signs of breakdown of the perturbative 
approach. Besides, the lifetime ratio $\tau(D^+) /\tau(D^0)$, which is insensitive to SU(3) breaking 
and not subject to the GIM suppression, agrees with the data \cite{PDG}, as calculated up to leading 
order in the $1/m_c$ expansion based on the formulation in \cite{Beneke:2002rj}. Therefore, it is the 
GIM cancellation in Eq.~(\ref{c12}) which strongly suppresses the 
perturbative contributions, and fails the inclusive analyses. It has been known that such cancellation 
does not take place in the $B_{s(d)}$ meson mixing, since the CKM factors in Eq.~(\ref{s12}) do not follow 
the pattern in the charm mixing. The CKM factors for the kaon mixing obey the similar pattern, 
$\lambda_{u}^{2}\approx-\lambda_u\lambda_c\approx\lambda_{c}^2$. However, only the first piece associated 
with $\lambda_{u}^{2}$ survives the phase space constraint, so the delicate cancellation does not happen 
either. Without the GIM cancellation, short-distance dynamics remains important in the $B_{s(d)}$ meson 
mixing and the kaon mixing.

\section{CONCLUSION}

We have analyzed the neutral meson mixing in the framework based on the dispersion relation, from which
the width difference of the two neutral meson mass eigenstates is solved directly.
The idea is to treat the dispersion relation as an inverse problem, in which 
nonperturbative observables at low mass are solved with perturbative inputs from high mass. 
It was emphasized that initial conditions of solutions at physical thresholds for involved
decay channels play an essential role. Their distinctions from the thresholds at the quark level
provide the nonperturbative effects, which determine how significantly the solution for each channel 
deviates from the corresponding perturbative input. The physical thresholds for various channels
induce the SU(3) symmetry breaking, which is
the key to explain the $D$ meson mixing. The threshold-dependent contributions, acting like nonperturbative 
power corrections in QCD sum rules, also stabilize the results of the mixing parameter $y(s=m_D^2)$ in the 
inverse problem: the convergence of the solutions in the polynomial expansion and the insensitivity to the 
arbitrary transition scale, which was introduced through the ultraviolet regularization of the dispersive 
integrals, have been demonstrated. In this sense, our formalism is free of tunable parameters, and this 
work represents an improvement of our previous one, which relies on a discretionary parametrization for 
the mixing parameter $y(s)$. 

It is intriguing to find that the solutions of $y(s)$ exhibit several oscillations, which reflect the
alternate opening of the destructive and constructive channels with the increase of the phase space.
The peak of the function $y(s)$ around the $D$ meson mass with the height greater than in the previous
exclusive analyses based only on two-body modes suggests that nearby resonances or multi-particle 
decays give the sizable contributions to $y(m_D^2)$. The channel with two 
strange quarks, i.e., di-kaon states, provides the major source of the SU(3) breaking relative to the 
channels with two down quarks and with one down quark and one strange quark, which enhances the net 
contribution to $y(m_D^2)$ by four orders of magnitude compared 
with the perturbative inputs. The mixing parameter $x(m_D^2)$ was derived from the
dispersive integration of $y(s)$, to which the contributions from the three channels containing $b$ quarks 
are negligible. The solutions for the various channels can be employed to calculate the mixing
parameters in both the $CP$-conserving and $CP$-violating cases: we simply multiply the solutions by the
associated CKM factors without and with the imaginary parts, respectively. It has been argued that our 
results for $x(m_D^2)$ and $y(m_D^2)$ can accommodate the data, after the enhancement from the matrix 
element of the $(S-P)(S-P)$ effective operator is taken into account. The theoretical uncertainty in 
our method is controllable, reflected by the very flat plateau of $y(m_D^2)$ in the stability window of 
$\Lambda$. In addition, we have predicted the coefficient ratio $q/p$ in the $CP$-violating case,  
which can be scrutinized by future precise measurements.

We have also studied the $B_{s(d)}$ meson mixing and the kaon mixing in the same framework.
It was found that the deviation of the solution from the corresponding perturbative input
is at the $O(10\%)$ level for each channel in the width difference, and the breakdown 
of the quark-hadron duality is similar to the amount in the charm mixing. Because there exists no
or milder cancellation of the perturbative pieces among the different channels, 
short-distance dynamics can be relatively important. Hence, the duality 
violation is not the major cause that renders the $D$ meson mixing special from the others.
It is the GIM cancellation that makes the tiny perturbative contributions in the inclusive analyses, 
in contrast to which the SU(3) breaking effects manifest in the $D$ meson mixing. We stress that our
work does not aim at a precise calculation and an exact match to the data, but at the verification that 
the box-diagram contributions can be greatly enhanced to the order of magnitude of the observed
charm mixing, and the neutral meson mixing, no matter whether it is governed by perturbative or 
nonperturbative dynamics, can be addressed consistently and systematically in our formalism. 


To improve the precision of the predictions, more accurate hadronic matrix elements of the effective 
operators, available higher-order corrections to the effective weak  Hamiltonian 
\cite{Grozin:2017uto,Gerlach:2022hoj}, 
subleading contributions from heavy quarks \cite{Brod:2022har}, and corrections with amplitudes being 
topologically distinct from the box diagrams, like the double penguin contribution \cite{Petrov:1997ch}, 
can be included into the inputs of our method. Simply speaking, the ultimate precision of the results
is controlled by the accuracy of the inputs at large mass, i.e., of our understanding on 
the $B$ meson mixing. It is then promising to lower the uncertainties down
to 10\% level \cite{Gerlach:2022hoj}. A thorough picture of the neutral meson mixing 
mechanism will help explorations of other observables, such as effects of 
the $D$ meson mixing in the extraction of the weak phase $\gamma$ from the $B\to D K$ decays 
\cite{Rama:2013voa,Harnew:2014zla}, and the determination of the quantity $y_{CP}$ from the 
$D\to K\pi$, $KK$ decays \cite{Schwartz:2022egt}. Once the $D$ meson mixing is realized, 
relevant data, such as those associated with the 
coefficient ratio $q/p$, can be used to constrain new physics models
\cite{Burdman:1995yy,Chivukula:2010tn,Faessler:2010az,Aranda:2010cy,Buras:2010zm,
Trott:2010iz,Adachi:2011tn,Nandi:2011uw,Descotes-Genon:2011rgs,Lee:2013esa,Cheng:2015lsa,
Hati:2015awg,Hu:2019heu,Buras:2021rdg,Oliveira:2022vjo,CarcamoHernandez:2022fvl}.
Our formalism is expected to have potential and broad applications in phenomenology.


\section*{Acknowledgement}

We are grateful for helpful discussions with A. Lenz, H. Umeeda, F.R. Xu and F.S. Yu.
This work was supported in part by National Science and Technology Council of the Republic
of China under Grant No. MOST-110-2811-M-001-540-MY3.


\end{document}